\documentclass[aps,twocolumn,amsmath,amssymb,floatfix]{revtex4-1}
\usepackage{graphicx}
\usepackage{dcolumn}
\usepackage{bm}
\usepackage{amsfonts}
\usepackage{graphicx}% Include figure files
\usepackage{dcolumn}% Align table columns on decimal point
\usepackage{bm}% bold math
\usepackage{amsmath}% needed for subequations
\usepackage{amssymb}

\begin{document}
\title{Interplay between phonon and impurity scattering in 2D hole transport}
\author{Hongki Min$^{1}$}
\author{E. H. Hwang$^{2}$}
\author{S. Das Sarma$^{2}$}
\affiliation{
$^{1}$Department of Physics and Astronomy, Seoul National University, Seoul 151-747, Korea\\
$^{2}$Condensed Matter Theory Center, Department of Physics, University of Maryland, College Park, Maryland 20742, USA
}
\date{\today}

\begin{abstract}
We investigate temperature dependent transport properties of
two-dimensional p-GaAs systems taking into account both hole-phonon and 
hole-impurity scattering effects. By analyzing the hole mobility data of
p-GaAs in the temperature range 10 K$<T<$100 K,
we estimate the value of the appropriate deformation potential for hole-phonon
coupling. 
%We find that the value of deformation potential coupling
%could change by as much as 60\% (i.e. $D=7.6-12.7$ eV) depending on the
%value of the depletion density $n_{\rm depl}$.
Due to the interplay between hole-phonon and hole-impurity scattering
the calculated temperature-dependent resistivity shows interesting nonmonotonic
behavior. In particular, we find that there is a temperature range
(typically 2 K$<T<$10 K) in 
which the calculated resistivity becomes independent of temperature
due to a subtle cancellation between the temperature dependent
resistive scattering contributions arising from impurities and phonons. This
resistivity saturation regime appears at low carrier densities when
the increasing resistivity 
due to phonon scattering compensates for the decreasing resistivity
due to the nondegeneracy effect. 
This temperature-independent flat resistivity regime is experimentally
accessible and may have already been observed in a recent experiment.
\end{abstract}

\maketitle
%\normalsize

\section{background}

The room-temperature resistivity of a metal (as well as most metallic
electronic materials, e.g., doped semiconductors) is mainly limited by
the electron-phonon interaction \cite{ziman} i.e., by phonon
scattering, with a notable exception being doped graphene with its
very weak electron-phonon coupling \cite{g_phonon1}. The
electron-phonon scattering contribution to the resistivity falls off
strongly at very low temperatures ($T < T_{\rm BG}$) in the so-called
Bloch-Gr\"{u}neissen (BG) regime \cite{ziman}, due to the exponential
suppression of the bosonic thermal occupancy of the phonons. At low
temperatures, therefore, the electronic conductivity of metallic
systems is invariably limited by disorder, i.e., by electron-impurity
scattering, which gives rise to the zero-temperature residual
resistivity of metals at low temperatures where the phonon scattering
contribution has vanished. In general, the impurity scattering
contribution to the metallic resistivity is temperature independent,
at least in three-dimensional (3D) metals, because the impurities are
quenched, and the temperature scale is, therefore, the Fermi
temperature ($T_{\rm F} \sim 10^4$ K in 3D metals), the impurity scattering
contribution to the resistivity is essentially temperature-independent
in the $0-300$ K regime. This, however, is not true if $T_{\rm F}$ is low as
it could be in 2D semiconductor-based systems with their tunable
carrier density where at low densities $T_{\rm F}$ could be just a few
K. In fact, for 2D GaAs hole systems (2DHS), the situation with $T_{\rm F}
<1$ K can easily be reached for a 2D hole density $\sim 10^{10}$
cm$^{-2}$ \cite{holeexp,mills1999}. This situation, which has no known analog in
3D metallic systems, leads to very strong experimentally observed
temperature-dependent 2DHS resistivity in the $T\alt 1$ K temperature
range arising entirely from electron-impurity scattering since the
phonon scattering is completely thermally quenched at such low
temperatures. As an aside, we note that the temperature dependence of
2D electronic resistivity arising from impurity scattering in 2D GaAs
based electron systems is rather weak due to the much lower effective
mass of 2D electrons ($m_{\rm e} \sim 0.07 m_0$, where $m_0$ is the vacuum
electron mass), leading to much higher $T_{\rm F}$, compared with 2D holes
($m_{\rm h} \sim 0.4 m_0$) in GaAs-based 2D heterostructures.

The goal of the current work is to explore the interplay between
impurity scattering and phonon scattering in the temperature-dependent
resistivity of 2DHS in GaAs-based 2D systems. At higher temperature
($\agt 100$ K), the GaAs carrier resistivity is completely dominated by
longitudinal optical (LO) phonon scattering, which has been extensively studied
\cite{kawamura1992} and is not a subject matter of interest here since the
resistivity limited by LO-phonon scattering manifests the strong
exponential temperature dependence $\sim e^{-\hbar\omega_{\rm LO}/k_{\rm B}T}$
with $\hbar\omega_{\rm LO} \sim 36$ meV.

Our interest is in the interplay between acoustic phonon scattering
and impurity scattering in the low to intermediate temperature regime
($T\sim T_{\rm BG} - T_{\rm F}$) where both impurity scattering and acoustic
phonon scattering contributions to resistivity would show nontrivial
temperature dependence. In particular, we are interested in the question
whether an interplay between the two scattering processes could lead
to an approximately constant (i.e. temperature independent)
resistivity over some intermediate temperature range (1 K $<T<$
40 K). One of our motivation comes from a recent experiment \cite{zhou2012}
which discovered such an intermediate-temperature ``resistivity
saturation'' phenomenon in a 2D GaAs-based electron system in the
presence of a parallel magnetic field (which
presumably serves to enhance the electron effective mass due to the
magneto-orbital coupling, thus reducing the Fermi temperature of the
electron system \cite{parallel_B}). A second motivation of our work is estimating the
deformation potential coupling strength for hole-acoustic phonon
scattering in 2DHS in GaAs. It turns out that the electron-phonon
deformation coupling is not known in GaAs, and a quantitative
comparison between our theoretical results and experimental transport
data could lead to an accurate estimation of the deformation potential
coupling in the GaAs-based 2DHS. We mention in this context that the
accurate evaluation of the electron deformation potential coupling in
2D GaAs systems is also based on a quantitative comparison of the
experimental and theoretical transport data \cite{kawamura1992,kawamura1990}.

The basic physics we are interested in (see Fig.~1) is a situation
where the acoustic phonon contribution to the resistivity increases linearly
with increasing temperature ($T>T_{\rm BG}$), but the impurity
contribution decreases with increasing temperature ($T \agt T_{\rm F}$) as
happens in a nondegenerate classical system where increasing
temperature must necessarily increase the conductivity since the
electrons are classically moving ``faster''. (We emphasize that such a
situation is physically impossible in 3D metals since $T>10^{4}$ K
would be required, but is routinely achieved in 2D semiconductor-based
systems where very low carrier density could lead to very low values
of $T_{\rm F}$.) We explicitly consider a situation with $T \ll T_{\rm LO}$, where
$T_{\rm LO} \sim 100$ K in GaAs where LO-phonon start contributing to the
resistivity in a substantial manner. Now we ask the question whether
it is possible for the increasing temperature dependent contribution
to the hole resistivity 
arising from enhanced phonon scattering at higher
temperatures could be approximately canceled over a finite temperature
range by the decreasing
temperature dependent contribution arising from impurity scattering
with increasing temperature due to the quantum-classical crossover,
i.e., the nondegeneracy, effects. Some early theoretical work
\cite{hwang2000} indicated that the interplay between phonon
scattering and nondegeneracy may indeed lead to a partial cancellation
of different temperature-dependent contributions, producing an
interesting nonmonotonicity (see Fig.~1) in the resistivity as a
function of temperature in the $1-10$ K range for low-density,
high-mobility 2DHS in GaAs. In the current work, we look into this
issue at great depth in view of the recent experimental work. In
addition, we obtain the appropriate 2DHS deformation potential
coupling constant by comparing our theoretical results to existing
2DHS hole transport data.

\section{Introduction}
\label{sec:motivation}

The temperature dependent transport in 2D systems
has been a subject of intense activity since the observation of an
apparent electronic metal-insulator transition (MIT), 
which represents the experimental observation of a transition from
an apparent metallic behavior (i.e., $d\rho/dT>0$, where $\rho$ is the
2D resistivity) to an insulating behaviors (i.e. $d\rho/dT <0$) as the
carrier density is reduced.
The remarkable  observation of the  anomalous  metallic
temperature dependence of the resistivity
is observed mostly in high-mobility low-density 2D semiconductor
systems \cite{dassarma2005,abrahams2001}.
The low temperature anomalous
metallic behavior discovered in 2D semiconductor systems
arises from the physical mechanism of strong temperature dependent
screening of charged impurity scattering \cite{dassarma2005}.
At low temperatures ($T \alt T_{\rm F}$, where $T_{\rm F}=E_{\rm
  F}/k_{\rm B}$ is the Fermi 
temperature with the Fermi energy $E_{\rm F}$) the main scattering mechanism 
in resistivity is due to impurity disorder from unintentional background
charged impurities and/or intentional dopants in the modulation-doping.
The resistivity $\rho_{\rm imp}(T)$ limited by the charged
impurities increases linearly with temperature at lower temperatures
($T<T_{\rm F}$) due to screening effects. 
This is a direct manifestation of the weakening of the screened
charged disorder with increasing temperature \cite{dassarma2005,dassarma1999} or
equivalently an electron-electron interaction effect in the so-called
ballistic regime \cite{zna}.
For $T \gg T_{\rm F}$, $\rho(T)$
decreases as $T_{\rm F}/T$ due to nondegeneracy effects and the
quantum-classical crossover occurs 
at the intermediate temperature regime around $T\sim T_{\rm F}$. \cite{dassarma1999}

The temperature dependent resistivity \cite{kawamura1992} limited by
acoustic phonons $\rho_{\rm ph}(T)$ undergoes a smooth 
transition from	a linear-$T$ dependence at high ($T>T_{\rm BG}$)
temperatures to a weak high power $T^{a}$
dependence with $a\geq 3$ as the temperature is reduced below
$k_{\rm B}T_{\rm BG} = 2k_{\rm F}v_{\rm ph}$, where $k_{\rm F}$ 
is the Fermi wave vector of the 2D hole system and $v_{\rm ph}$ is
the longitudinal or transversal sound velocity. 
The characteristic temperature $T_{\rm BG}$ is referred to as the
Bloch-Gr\"{u}neisen (BG) temperature. 
Note that $T_{\rm BG}$ is much smaller than the Debye temperature
since the inverse lattice constant greatly exceeds $k_{\rm F}$. 
In the BG regime ($T<T_{\rm BG}$) the scattering rate is strongly reduced by the
thermal occupation factors because the phonon population 
decreases exponentially and the phonon emission is prohibited by the sharp
Fermi distribution, which gives rise to high power law behavior
($a\geq 5$ with screening effects, but $a\geq 3$ without screening) in
the temperature 
dependent resistivity. 
Thus, the phonon contribution to the resistivity is negligible
compared to the charged impurities 
and the phonon contribution to the resistivity shows very weak
temperature dependence for $T<T_{\rm BG}$. 
For temperatures $T>T_{\rm BG}$, since the electron-phonon scattering becomes
proportional to the square of the oscillation amplitude of ions,
the $\rho_{\rm ph}$ depends linearly on the temperature. 
We note that at low carrier density where $k_{\rm F}$ is small, $T_{\rm BG}$ can
be very low.

In this paper we investigate the temperature dependent transport
properties of p-type GaAs-based 2DHSs for the temperatures
$T \alt 100$ K by taking into account both hole-phonon and
hole-impurity scatterings. In p-GaAs, holes interact with
acoustic phonons through a short-range deformation potential as well
as through a long range electrostatic potential resulting from the
piezoelectric effect. The precise value of the deformation-potential
coupling constant $D$ is very important to understand 2D p-GaAs
carrier transport properties. For example, the mobility is strongly related
to the deformation potential, $\mu \sim D^{-2}$. However,
the precise value of $D$ for p-GaAs has not been available, so the
value of the n-GaAs deformation potential ($D=12$ eV
\cite{kawamura1992}) is used uncritically. 
Thus, the current investigation of the transport properties of p-GaAs
systems is motivated by getting the exact value of the deformation potential
constant $D$ in p-GaAs. 
%We obtain $D=9$ eV as the most suitable value for the p-GaAs acoustic phonon deformation coupling constant by fitting the available experimental data. The value of $D=9$ eV is larger than the generally accepted value in bulk GaAs ($D=7$ eV) \cite{wolfe1970} but smaller than $12$ eV of the n-GaAs.
We find that the fitted value of deformation-potential coupling
could change by as much as 60\% (i.e. $D=7.6-12.7$ eV) depending on the
value of the depletion density $n_{\rm depl}$.
When we assume $n_{\rm depl}=0$, we obtain $D=12.7$ eV as the most
suitable value for the p-GaAs acoustic phonon deformation coupling
constant by fitting the available experimental data \cite{noh2003}. 
The value of
$D=12.7$ eV is larger than the generally accepted value in bulk GaAs
($D=7$ eV) \cite{wolfe1970} but comparable to the value of the
n-GaAs ($12-14$ eV) \cite{price1985,mendez1984,kawamura1990}. 
However, due to the lack of knowledge about $n_{\rm depl}$ we have
the uncertainty in the value of deformation potential coupling.
To precisely determine the value of $D$ for holes it is required a careful
measurement of $\rho(T)$ over the $T=1-50$ K range in a high mobility and
high density ($n \agt 5\times10^{11}$ cm$^{-2}$) sample.
Note that our finding of the apparent dependence of the GaAs 2D hole
deformation potential coupling on the background (and generally
unknown) depletion charge density in the 2DHS is obviously {\it not} a
real effect since the hole-phonon coupling strength cannot depend on
the depletion charge density. The apparent dependence we find arises
from the fact that the calculated resistivity depends strongly on the
depletion density whose precise value is unknown.

%%%%%%%%%%%%%%%%%%%%%%%%%%%%%%%%%

\begin{figure}%[htb]
\includegraphics[width=\linewidth]{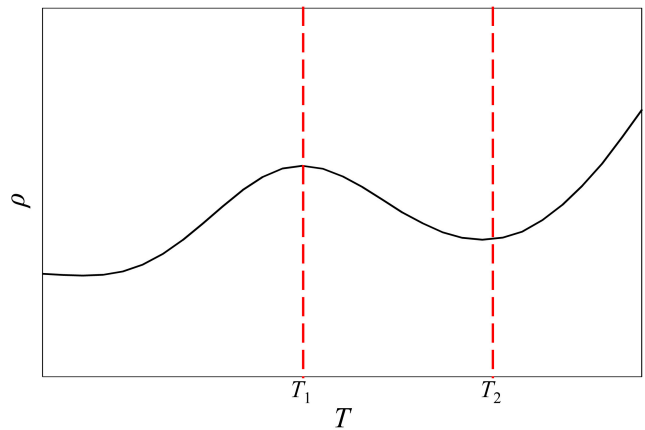}
\caption{(Color online) (a) Schematic resistivity behavior in the
  presence of both hole-impurity 
  scattering and acoustic phonon scattering as a function of
  temperature. Typically $T_1\sim T_{\rm F}$ and $T_2\sim 5-10$ K depending
  on the carrier density. } 
\label{fig:rho_T_schematic}
\end{figure}

Other motivation of our work is to investigate the nontrivial nonmonotonic
behavior of the temperature dependent resistivity observed experimentally.
In the presence of the hole-phonon and hole-impurity scattering 
a nonmonotonicity arises from a competition among three
mechanisms \cite{mills1999,hwang2000,lilly2003}: screening which is 
particularly important for $T<T_{\rm F}$, nondegeneracy and the associated
quantum-classical crossover for $T \sim T_{\rm F}$, and 
phonon scattering effect which becomes increasingly important for $T >5-
10$ K, depending on the carrier density. 
In Fig.~\ref{fig:rho_T_schematic} we show the
schematic resistivity behavior of the p-GaAs in the presence of both charged
impurity and  phonon scatterings.
At low temperatures ($T<T_1\sim T_{\rm F}$),  the scattering arising from charged
impurities dominates and the resistivity increases as temperature increases 
due to the screening effects.
For $T > T_2\sim 5-10$ K the scattering by acoustic
phonons plays a major role and limits the carrier mobility of p-GaAs
systems in this temperature range. Note that in general $T_2 \gg
T_{BG}$ for p-GaAs systems. However, at intermediate
temperatures ($T_1<T<T_2$) the 
competition between acoustic phonon scattering and impurity scattering
gives rise to a nontrivial transport behavior.
We carefully study the non-trivial transport properties of p-GaAs in the
intermediate temperature range (i.e., $T_1<T<T_2$). Interestingly we
find that the approximate temperature dependent $\rho(T)$ in an intermediate
temperature range can be constant, which arises from the approximate
cancellation between the quantum-classical crossover and phonon
scattering, and is quite general.  
When the phonon scattering dominates impurity scattering for $T>T_2$ the
temperature dependence of hole mobility enables one to extract 
information on the electron-phonon scattering from mobility
measurements. In this temperature range the valence band deformation
potential can be determined by fitting theoretical calculations to
the existing carrier mobility data. In this paper 
we extract the value of the deformation potential by fitting the experimental
mobility data, but uncertainty arises from the unknown values of the
depletion charge density and background impurity charge density.

%%%%%%%%%%%%%%%%%%% Fig. 1  %%%%%%%%%%%%%%%%%%%%%%%%%
%%%%%%%%%%%%%%%%%%%%%%%%%%%%%%%%%%%%

To investigate the temperature dependent transport properties for the
p-GaAs 2DHS we use the finite temperature Boltzmann
transport theory considering scatterings by charged random impurity centers and
by electron-acoustic phonons with finite temperature and finite wave
vector screening through random phase approximation (RPA) \cite{dassarma2005}. 
We also include the finite size confinement
effects (i.e., we take into account the extent of the 2D system in the
third dimension and do not assume it to be a zero-width 2D layer).
The effect we neglect in our theory is the inelastic optical
phonon scattering. Polar carrier scattering by LO phonons is 
important in GaAs only at relatively high temperatures ($T \agt
100$ K), becoming dominant at room temperatures.  Due to the rather
high energy of the GaAs optical phonons ($\sim 36$ meV), LO phonon
scattering is completely suppressed in the temperature regime
($T<100$ K) of interest to us in this work. Resistive scattering by
optical phonons in 2D  GaAs system has been considered in the
literature \cite{kawamura1992}. We note that the other scattering
mechanisms (e.g., interface roughness scattering and alloy disorder scattering in
Ga$_x$Al$_{1-x}$As, etc.) are known to be much less quantitatively
important \cite{ando1982} than the mechanisms we are considering (i.e., impurity scattering and acoustic
phonons) in this work.

%%%%%%%%%%%%%%%%%%%%%%%%%%%%%%%%%%%%%%%%%%%%%%%%%%

The rest of the paper is organized as follows. In Sec. III we present
the general theory of the impurity and phonon scatterings, and discuss the power-law behavior 
of the resistivity in the low and high temperature limit. 
In Sec. IV we show our calculated resistivity taking into account hole-phonon and hole-impurity scatterings, 
and demonstrate that there is a temperature-independent region due to the competition between the two scatterings.
Finally we conclude in Sec. V summarizing our results.

%%%%%%%%%%%%%%%%%%%%%%%%%%%%%%%%%%%%%%%
 
\section{Theory}
\label{sec:theory}

To investigate the temperature dependent resistivity $\rho(T)$ (or
equivalently conductivity $\sigma(T) \equiv \rho(T)^{-1}$) of p-GaAs
systems we start with the Drude-Boltzmann semiclassical formula for 2D
transport. Due to the finite extent in the $z$-direction of the real 2D
semiconductor system we have to include a form factor depending on the
details of the 2D structure. 
In GaAs heterostructures, the carriers are spatially confined at the
2D interface and there is no longer translational
invariance along the direction normal to the interface, designated as
the $z$ direction. We assume that the confinement profile is described
by the variational wavefunction $\Psi(x,y,z)=\xi_0(z)e^{i (k_x x+k_y
  y)}$ \cite{ando1982}, where  
\begin{equation}
\label{eq:WF_z}
\xi_0(z)=\sqrt{{1\over 2} b^3 z^2} \exp\left(-{1\over 2} bz\right),
\end{equation}
and $b$ is a variational parameter. For a triangular well, $b$ is
given by \cite{fang1966} 
\begin{equation}
b=\left({48\pi m^{\ast} e^2 \over \epsilon_0 \hbar^2}\right)^{1\over
  3}\left(n_{\rm depl}+{11\over 32}n\right)^{1\over 3}, 
\end{equation}
where $m^{\ast}$ is an effective mass, $\epsilon_0$ is the dielectric constant,
$n_{\rm depl}$ is the depletion charge per unit area and $n$ is the 2D carrier density. 

The density of states of 2DHS with
parabolic energy dispersion $\varepsilon({\bm k})={\hbar^2 k^2 \over 2
  m^{\ast}}$ is given by 
\begin{equation}
\label{eq:DOS}
D(\varepsilon)=
\begin{cases}
D_0 & \text{($\varepsilon > 0$)}, \\
0 & \text{($\varepsilon < 0$)},
\end{cases}
\end{equation}
where $D_0={g_{\rm s} m^{\ast} \over 2\pi \hbar^2}$ and $g_{\rm s}=2$ is the spin degeneracy factor. 
The carrier density $n$ is given by
\begin{equation}
\label{eq:n}
n=D_0 \int_0^{\infty} d\varepsilon f(\varepsilon)=D_0 k_{\rm B} T
\ln\left[1+\exp\left({\mu(T) \over k_{\rm B} T}\right)\right], 
\end{equation}
where $f(\varepsilon)=[e^{\beta (\varepsilon-\mu)}+1]^{-1}$ is the
Fermi distribution function and $\beta=1/(k_{\rm B} T)$. 
Alternatively, the chemical potential $\mu(T)$ at a finite temperature
$T$ can be expressed as 
\begin{equation}
\label{eq:mu}
\mu(T)=k_{\rm B} T \ln\left[\exp\left({E_{\rm F}/k_{\rm B} T}\right)-1\right],
\end{equation}
where $E_{\rm F}=n/D_0$. Note that $\lim_{T\rightarrow 0}
\mu(T)=E_{\rm F}$. 
From Eqs.~(\ref{eq:n}) and (\ref{eq:mu}),
\begin{equation}
{\partial n \over \partial \mu}=D_0
\left[1-\exp\left(-{E_{\rm F}\over k_{\rm B}
    T}\right)\right]. 
\end{equation}

The finite temperature Thomas-Fermi (TF) screening wavevector
$q_{\rm TF}(T)$ is defined by  
\begin{equation}
\label{eq:TF_T}
q_{\rm TF}(T)={2\pi e^2 \over \epsilon_0} {\partial n \over \partial
  \mu}= q_{\rm TF} \left[1-\exp\left(-{E_{\rm F}\over k_{\rm
      B} T}\right)\right], 
\end{equation}
where $q_{\rm TF}={2\pi e^2 \over \epsilon_0} D_0={g_{\rm s}
  e^2 m^{\ast} \over \epsilon_0 \hbar^2}$. 
Note that
\begin{equation}
\label{eq:q_TF_limit}
q_{\rm TF}(T) =
\begin{cases}
q_{\rm TF} & \text{($T\rightarrow 0$)}, \\
q_{\rm TF} \left({E_{\rm F} \over  k_{\rm B} T}\right) &
\text{($T \rightarrow \infty$)}. 
\end{cases}
\end{equation}

\subsection{Electron-phonon interactions}
The electron-longitudinal acoustic phonon interaction Hamiltonian for
a small phonon wavevector ${\bm Q}$ is  
\begin{equation}
H_{\rm DP}({\bm Q})=D {\bm Q}\cdot{\delta{\bm R}},
\end{equation}
where $D$ is the acoustic phonon deformation potential and $\delta {\bm R}$ is the displacement vector. 
Note that position operator in a simple harmonic oscillator with mass
$m$ and angular frequency $\omega$ is given by %\cite{sakurai1994} 
\begin{equation}
x=\sqrt{\hbar \over 2 m \omega} (a + a^{\dagger}),
\end{equation}
where $a$ and $a^{\dagger}$ is the annihilation and creation operators, respectively.
Similarly, $\delta{\bm R}$ for $\bm Q$ can be expressed in terms of phonon
annihilation and creation operators as 
%\begin{equation}
%\delta{\bm R}({\bm x})=\int {d^3 Q \over (2\pi)^3} \sqrt{\hbar \over 2 \rho_{\rm m} \omega_{\bm Q}} \hat{e}_{\bm Q} \left( a_{\bm Q} e^{i{\bm Q}\cdot {\bm x}}+ a_{\bm Q}^{\dagger} e^{-i{\bm Q}\cdot {\bm x}}\right) 
%\end{equation}
%or in a momentum space,
\begin{equation}
\delta{\bm R}({\bm Q})= \sqrt{\hbar \over 2 \rho_{\rm m} \omega_{\bm Q}} \hat{e}_{\bm Q} \left( a_{\bm Q} + a_{-\bm Q}^{\dagger}
\right), 
\label{eq:electron_phonon}
\end{equation}
where $\rho_{\rm m}$ is the mass density, $\omega_{\bm Q}=v_{\rm l}
Q$, $v_{\rm l}$ is the longitudinal sound velocity and $\hat{e}_{\bm Q}$ is the phonon polarization unit vector. 
Thus the electron-phonon interaction by the deformation potential coupling
can be expressed as
\begin{equation}
H_{\rm ep}=D \sum_{\bm Q} \sqrt{\hbar \over 2 \rho_{\rm m} \omega_{\bm
    Q}} {\bm Q} \left( a_{\bm Q} + a_{-\bm Q}^{\dagger} \right)  \rho({\bm Q}) 
\end{equation}
where $\rho({\bm Q})$ is the electron density operator.

For the piezoelectric
scattering in polar semiconductors (i.e. GaAs) the scattering matrix
elements are equivalent to the substitution of the deformation with the
following form \cite{zook1964}
\begin{equation}
%D^2 \rightarrow \frac{(eh_{14})^2 A}{q^2+q_z^2}
D^2 \rightarrow \frac{(eh_{14})^2 A}{Q^2}
\end{equation}
where $h_{14}$ is the basic piezoelectric tensor component and $A$ is
a dimensionless anisotropy factor that depends on the direction of
the phonon wave vector in the crystal lattice. We provide details of
the parameter $A$ in the following subsection.

\subsection{Boltzmann transport theory}

We calculate the temperature dependence of the hole resistivity by
considering screened acoustic-phonon scattering. We include both
deformation potential and piezo-electric coupling of the 2D holes to
3D acoustic phonons of GaAs. Details of the acoustic-phonon scattering
theory are given in Ref.~\cite{kawamura1992}.

The transport relaxation time $\tau(\varepsilon_{\bm k})$ at an energy $\varepsilon_{\bm k}$ and a 2D wavevector $\bm k$ is given by
\begin{equation}
\label{eq:tau_IE}
{1\over \tau(\varepsilon_{\bm k})}=\int {d^2 k' \over (2\pi)^2} W_{{\bm k},{\bm k}'} (1-\cos\phi_{{\bm k}{\bm k}'}) {1-f(\varepsilon_{{\bm k}'}) \over 1-f(\varepsilon_{\bm k})},
\end{equation}
where $W_{{\bm k},{\bm k}'}$ is the scattering probability between ${\bm k}$ and ${\bm k}'$ states, and $\phi_{{\bm k}{\bm k}'}$ is the scattering angle between ${\bm k}$ and ${\bm k}'$ vectors.

First, consider impurity scattering. Assume that impurity charges are
distributed  randomly in a 2D plane located at $(-d_{\rm imp})\hat{\bm z}$ with a 2D impurity density $n_{\rm imp}$. Then the effective impurity potential for a 2D wavevector $\bm q$ is given by
\begin{eqnarray}
V_{\rm imp}({\bm q},d)&=&\int_0^{\infty}dz \xi_0^2(z) \left({2\pi e^2 \over \epsilon_0 q} e^{-q(d_{\rm imp}+z)}\right) \nonumber \\
&=&\left({2\pi e^2 \over \epsilon_0 q} e^{-q d_{\rm imp}}\right)\left({b\over b+q}\right)^3.
\end{eqnarray}
From the Fermi's golden rule, $W_{{\bm k},{\bm k}'}$ for the impurity scattering has the following form:
\begin{equation}
W_{{\bm k},{\bm k}'}^{\rm imp}={2\pi \over \hbar} {|V_{\rm imp}({\bm q},d)|^2 \over \epsilon^2({\bm q},T) } n_{\rm imp} \delta (\varepsilon_{\bm k}-\varepsilon_{\bm k}'),
\end{equation}
where ${\bm q}={\bm k}-{\bm k}'$.
The dielectric function $\epsilon({\bm q},T)$ takes into account the screening effects of electron gas at a wavevector ${\bm q}$ and a temperature $T$. We will consider screened scattering within RPA approximation defined by 
\begin{equation}
\label{eq:screen_q}
\epsilon({\bm q},T)=1+{q_{\rm s}(q,T)\over q},
\end{equation}
where $q_{\rm s}(q,T)$ is the temperature- and wave vector-dependent
screening wavevector \cite{dassarma2011}. 
In the long wavelength limit ($q=0$), 
$q_{\rm s}(q,T)$ is given by Eq.~(\ref{eq:TF_T}). 

For electron-phonon scattering, $W_{{\bm k},{\bm k}'}$ for the electron-longitudinal acoustic phonon interaction has the following form \cite{ziman}:
\begin{equation}
W_{{\bm k},{\bm k}'}^{\rm ph}={2\pi \over \hbar} \int {dq_z \over 2\pi} {|C({\bm q},q_z)|^2 \over \epsilon^2({\bm q},T) } \Delta (\varepsilon_{\bm k},\varepsilon_{\bm k}') \left|I(q_z)\right|^2,
\end{equation}
where $I_z(q_z)$ is the wavefunction overlap at $q_z$ defined by
\begin{equation}
I_z(q_z)=\int dz \xi_0^2(z) e^{iq_z z}.
\end{equation}
Note that from Eq.~(\ref{eq:WF_z}), $\left|I_z(q_z)\right|^2={b^6 \over (b^2+q_z^2)^2}$.

The factor $C({\bm q},q_z)$ is the matrix element for acoustic phonon scattering.
From Eq.~(\ref{eq:electron_phonon}), for the deformation potential (DP),
\begin{equation}
\label{eq:Cdp}
|C_{\rm DP}({\bm q},q_z)|^2={D^2 \hbar Q \over 2\rho_{\rm m} v_{\rm l}},
\end{equation}
while for the piezoelectric potential (PE) \cite{kawamura1992},
\begin{equation}
\label{eq:Cpe}
|C_{{\rm PE},\lambda}({\bm q},q_z)|^2={(eh_{14})^2 \hbar A_{\lambda}(q,q_z) \over 2\rho_{\rm m} v_{\lambda} Q},
\end{equation}
where ${\bm Q}=({\bm q},q_z)$, $v_{\rm l}$ ($v_{\rm t}$) is the longitudinal (transverse) sound velocity and
\begin{equation}
A_{\rm l}(q,q_z)={9 q_z^2 q^4 \over 2(q_z^2+q^2)^3}, \,
A_{\rm t}(q,q_z)={8 q_z^4 q^2+q^6 \over 4(q_z^2+q^2)^3}.
\end{equation}

The factor $\Delta(\varepsilon,\varepsilon')$ is given by
\begin{equation}
\label{eq:Delta}
\Delta(\varepsilon,\varepsilon')=N_q \delta(\varepsilon-\varepsilon'+\hbar\omega_q)+(N_q+1)\delta(\varepsilon-\varepsilon'-\hbar \omega_q),
\end{equation}
where $N_q=(e^{\beta \hbar \omega_q}-1)^{-1}$ is the phonon occupation number. 
Note that the first and second terms in Eq.~(\ref{eq:Delta}) correspond to absorption and emission of phonons, respectively. 

Finally, the total transport relaxation time is given by
\begin{equation}
\label{eq:relaxation_time}
{1 \over \tau_{\rm tot}(\varepsilon)}={1 \over \tau_{\rm imp}(\varepsilon)}+{1 \over \tau_{\rm DP}(\varepsilon)}+{1 \over \tau_{\rm PE,l}(\varepsilon)}+{2 \over \tau_{\rm PE,t}(\varepsilon)},
\end{equation}
where in the last term the degeneracy of the transverse modes has been taken into account.
Then electrical conductivity in a 2DHS is given by
\begin{equation}
\label{eq:boltzmann}
\sigma= g_{\rm s} e^2 \int {d^2 k \over (2\pi)^2} {v^2_{\bm k} \over 2} \tau_{\rm tot}(\varepsilon_{\bm k}) \left(-{\partial f \over \partial \varepsilon}\right)_{\varepsilon=\varepsilon_{\bm k}},
\end{equation}
where $v_{\bm k}$ is the mean-velocity at ${\bm k}$. By inverting the
conductivity we have the resistivity, i.e. $\rho(T) = \sigma^{-1}(T)$.

\subsection{Quasi-elastic limit}
For a degenerate system ($k_{\rm B} T \ll E_{\rm F}$), all the scattering events at an energy $\varepsilon$ take place in a thin shell around the energy circle and the scattering can be considered as quasi-elastic. For electron-phonon scatterings, the transport relaxation time in Eq.~(\ref{eq:tau_IE}) is given by 
\begin{equation}
\label{eq:tau_QE}
{1\over \tau(\varepsilon)}={2\pi \over \hbar} D_0 \int {d\phi \over 2\pi} \int {dq_z \over 2\pi} {|C(q)|^2 \over \epsilon^2(q)} \left|I(q_z)\right|^2 G(q) (1-\cos\phi), 
\end{equation}
where $q=2 k_{\rm F} \sin (\phi/2)$ and $G(q)$ is given by
\begin{eqnarray}
\label{eq:G}
G(q)&=&\beta \int d\varepsilon f(\varepsilon) \left[\left(1-f(\varepsilon+\hbar\omega_q)\right) N_q\right. \\
&+& \left.\left(1-f(\varepsilon-\hbar\omega_q)\right) (N_q+1)\right] \nonumber \\
&=&2\beta \hbar \omega_q N_q (N_q+1). \nonumber
\end{eqnarray}

Thus for DP, Eq.~(\ref{eq:tau_QE}) becomes \cite{kawamura1990,kawamura1992}
\begin{eqnarray}
\label{eq:tau_DP}
{1\over \tau_{\rm DP}(\varepsilon)}&=&{3 D^2 m^{\ast} b  k_{\rm B} T\over 16 \hbar^3 \rho_{\rm m} v_l^2} \int_0^{\pi}{d\phi \over \pi} {(1-\cos\phi)\over \epsilon^2(q,T)} \nonumber \\
&\times&(\beta \hbar w_q)^2 N_q (N_q+1),
\end{eqnarray}
while for PE,
\begin{eqnarray}
\label{eq:tau_PE}
{1\over \tau_{{\rm PE},\lambda}(\varepsilon)}&=&{c_{\lambda} (e h_{14})^2 m^{\ast} k_{\rm B} T \over 2 \hbar^3 \rho_{\rm m} v_{\lambda}^2} \int_0^{\pi}{d\phi \over \pi} {(1-\cos\phi)\over q\epsilon^2(q,T)} \nonumber \\
&\times&(\beta\hbar w_q)^2 N_q (N_q+1) f_{\lambda}(q/b), 
\end{eqnarray}
where $c_{\rm l}=9/32$, $c_{\rm t}=13/32$ and
\begin{eqnarray}
f_{\rm l}(w)&=&{1+6w+13w^2+2w^3 \over (1+w)^6}, \\
f_{\rm t}(w)&=&{13+78w+72w^2+82w^3+36w^4+6w^5 \over 13(1+w)^6}. \nonumber
\end{eqnarray}

For simplicity, consider $f_{\lambda}(q/b)=1$ case assuming $b\gg 1$ in the extreme 2D limit.
Then, in the high temperature limit,
\begin{eqnarray}
{1\over \tau_{\rm DP}(T)}&\approx& {3 D^2 m^{\ast} b  k_{\rm B} T \over 16 \hbar^3 \rho_{\rm m} v_l^2} \propto T, \\
{1\over \tau_{\rm PE,\lambda}(T)}&\approx& {c_{\lambda} (e h_{14})^2 m^{\ast} \over 2 \hbar^3 \rho_{\rm m} v_{\lambda}^2} {4 \over \pi} {T \over T_{\rm BG} } \propto T, \nonumber
\end{eqnarray}
while in the low temperature limit,
\begin{eqnarray}
{1\over \tau_{\rm DP}(T)}&\approx& {3 D^2 m^{\ast} b  k_{\rm B} T
  \over 16 \hbar^3 \rho_{\rm m} v_l^2} {4\cdot 6! \zeta(6) \over \pi
  x_{\rm TF}^2} \left(T\over T_{\rm BG}\right)^5 \propto T^6,
\nonumber \\
{1\over \tau_{\rm PE,\lambda}(T)}&\approx& {c_{\lambda} (e h_{14})^2 m^{\ast} \over 2 \hbar^3 \rho_{\rm m} v_{\lambda}^2} {4\cdot 5! \zeta(5) \over \pi x_{\rm TF}^2} \left(T\over T_{\rm BG}\right)^5 \propto T^5, 
\end{eqnarray}
where $x_{\rm TF}=q_{\rm TF}/(2 k_{\rm F})$.
Note that the resistivity is proportional to the inverse relaxation time, thus it follows the same power-law dependence in the high or low temperature limit.

For the charged impurity with temperature-dependent screening wave vector
the asymptotic low- and high- temperature behaviors of 2D resistivity
for a $\delta$-layer system are given
by \cite{dassarma2004}
\begin{equation}
\rho_{\rm imp}(T\ll T_{\rm F}) \sim \rho_0 \left [ 1 +
  \frac{2 x_{\rm TF}}{1+x_{\rm TF}}\frac{T}{T_{\rm F}} + C\left ( \frac{T}{T_{\rm F}}
  \right )^{3/2} \right ], 
\end{equation}
\begin{equation}
\rho_{\rm imp}(T \gg T_{\rm F} ) \sim \rho_1 \frac{T_{\rm F}}{T} \left
[ 1 - \frac{3 \sqrt{\pi} x_{\rm TF}}{4} \left (\frac{T_{\rm F}}{T} \right )^{3/2}
  \right ],
\end{equation} 
where $\rho_0=\rho(T=0)$, $\rho_1=(h/e^2)(n_{\rm imp}/n \pi x_{\rm TF}^2)$, 
and $C= 2.646[x_{\rm TF}/(1+x_{\rm TF})]^2$.
At low temperatures ($T<T_{\rm F}$) the resistivity increases
linearly due to screening (or electron-electron interaction) effects
on the impurity scattering \cite{dassarma2005,dassarma1999}. To get
the linear temperature 
dependent resistivity at low temperatures 
it is crucial to include the temperature-dependent screening wave
vector \cite{dassarma2005}. At high temperatures ($T>T_{\rm F}$) $\rho(T)$
decreases inverse linearly due to nondegeneracy effects. Thus, it is
expected that
the resistivity has a maximum value and
the quantum-classical crossover occurs at the intermediate
temperature regime around $T_{\rm F}$.
When we consider both hole-phonon and hole-impurity scatterings 
the temperature dependent resistivity becomes nontrivial due to the
competition between these independent two scattering mechanisms.
Since the resistivity limited by charged impurities decreases at high
temperatures phonon scattering eventually takes
over and $\rho(T)$ increases with $T$ again, which gives rise to
nonmonotonicity in $\rho(T)$. The nonmonotonicity becomes
pronounced in the systems with strong impurity scattering or at low
carrier density. For weaker impurity scattering the phonon
scattering dominates before the quantum-classical crossover occurs, so
the overall resistivity increases with temperature. 
At higher carrier densities, $T_{\rm F}$ is pushed up to the phonon scattering regime,
and the quantum-classical cross-over physics is overshadowed by phonons
so that nonmonotonicity effects are not manifest.

%%%%%%%%%%%%%%%%%%%%%%%%%%%%%%%%%%%%%%%%%%%%%%%%%%%%

\section{Numerical results}
\label{sec:numerical}

\subsection{Determination of deformation potential}

\begin{figure}%[htb]
\includegraphics[width=\linewidth]{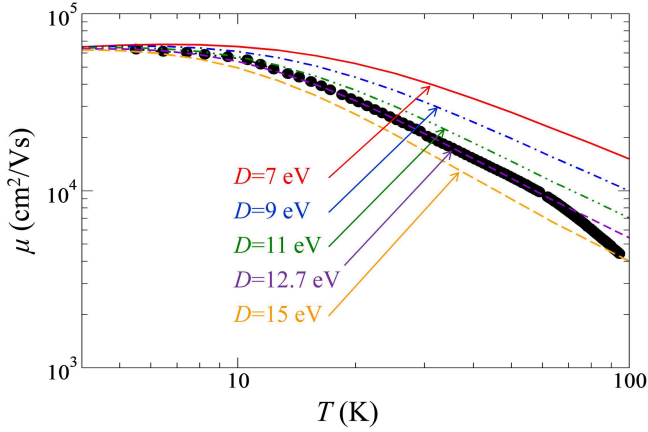}
\caption{(color online). Mobility as a function of temperature for several values of
  $D$ with $n=2\times 10^{11}$ cm$^{-2}$, $n_{\rm imp}=1.22\times
  10^{10}$ cm$^{-2}$, $n_{\rm depl}=0$ and $d_{\rm imp}=0$. Black dots
  represent the experimental data \cite{noh2003}.} 
\label{fig:fitting_D}
\end{figure}

In this section we provide the numerically calculated temperature
dependence of the hole resistivity by 
considering both screened acoustic-phonon scattering and screened
charged impurity scattering. 
In the calculation of phonon scattering we use the parameters corresponding to GaAs: 
$m^{\ast}=0.38$ $m_{\rm e}$, $v_{\rm l}=5.14\times 10^5$ cm/s,
$v_{\rm t}=3.04\times 10^5$ cm/s, $\rho_{\rm m}=5.3$ g/cm$^3$ and $e h_{14}=1.2\times 10^7$ eV/cm. 
For the deformation potential, we fitted several available mobility data set
\cite{stormer1984,noh2003} and the best fitted value we obtained is
$D=12.7$ eV for $n_{\rm depl}= 0$ (see Fig.~\ref{fig:fitting_D}). In the following calculations
we use this value as a deformation potential of p-GaAs.

%%%%%%%%%%%%%%%%%%%%%%%  Fig. 2%%%%%%%%%%%%%%%%%%%%%

%%%%%%%%%%%%%%%%%%%% Fig. 3 %%%%%%%%%%%%%%%%%%%%%%%%%%%%%%%%%%
\begin{figure}%[htb]
\includegraphics[width=\linewidth]{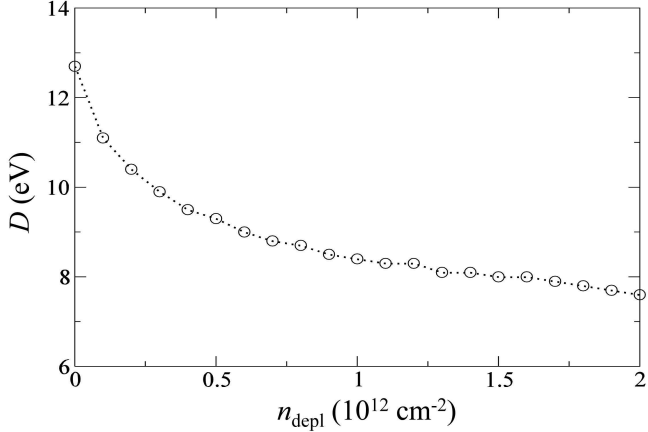}
\caption{Fitted deformation potential $D$ as a function of depletion density $n_{\rm depl}$ for a sample with $n=2\times 10^{11}$ cm$^{-2}$ and $d_{\rm imp}=0$.} 
\label{fig:D_ndepl}
\end{figure}

To obtain the best fitted value of the deformation potential, in
Fig.~\ref{fig:fitting_D}, we calculate the total mobility $\mu =
\sigma/ne$ with Eqs.~(\ref{eq:relaxation_time}) and (\ref{eq:boltzmann}) as a
function temperature for different values of the deformation-potential
constant $D$. Knowing the precise value of the deformation-potential coupling
constant $D$ is very critical because $\mu \sim D^{-2}$, i.e.,
the calculated mobility will be uncertain by a factor of four 
with values of $D$ differing by a factor of 2.
In this calculation we consider two different scattering
mechanisms: remote impurity scattering and acoustic phonon
scattering. We first fit the low temperature data ($T \alt 4$ K) to find the
charged impurity density, $n_{\rm imp}$, because the phonon scattering
is severely
suppressed and the charged impurity scattering determines the carrier
mobility in this temperature range. 
We set $d_{\rm imp}=0$ for simplicity and carried the effect of impurity by $n_{\rm imp}$.
For $n_{\rm depl}=0$, we find that $n_{\rm imp}=1.22\times
10^{10}$ cm$^{-2}$ gives the best fitted mobility at low temperatures for the data set. 
With this impurity density we calculate the mobility data at
high temperatures (20 K$<T<$60 K) by changing deformation potential.   
From Fig.~\ref{fig:fitting_D}, we get $D=12.7$ eV as the most suitable value for the p-GaAs 
acoustic phonon deformation coupling constant.

In Fig.~\ref{fig:D_ndepl} we show the deformation potential coupling
as a function of the depletion density for a hole density $n=2 \times
10^{11}$ cm$^{-2}$.
The calculated deformation potential $D$ strongly depends on the
depletion density $n_{\rm depl}$ for $n_{\rm depl} <10^{12}$ cm$^{-2}$,
but for higher densities ($n_{\rm depl} > 10^{12}$ cm$^{-2}$ it
decreases slowly, as seen in Fig.~\ref{fig:D_ndepl}.  
$n_{\rm depl}$ is a measure of fixed charges in the background and
typically $n_{\rm depl}$ is unknown. Thus, the uncertainty in the
value of deformation potential coupling (both for electrons and for
holes) could be a result of our lack of knowledge about $n_{\rm
  depl}$. When $n_{\rm depl}=0$ we get $D=12.7$ eV.
We use $D=12.7$ eV in the remaining calculations with $n_{\rm depl}=0$. However,  the
different values of $D$ do not change the results qualitatively.

%%%%%%%%%%%%%%%%%%%%%%%  Fig. 4 %%%%%%%%%%%%%%%%%%%%%%%%%%
\begin{figure}%[htb]
\includegraphics[width=\linewidth]{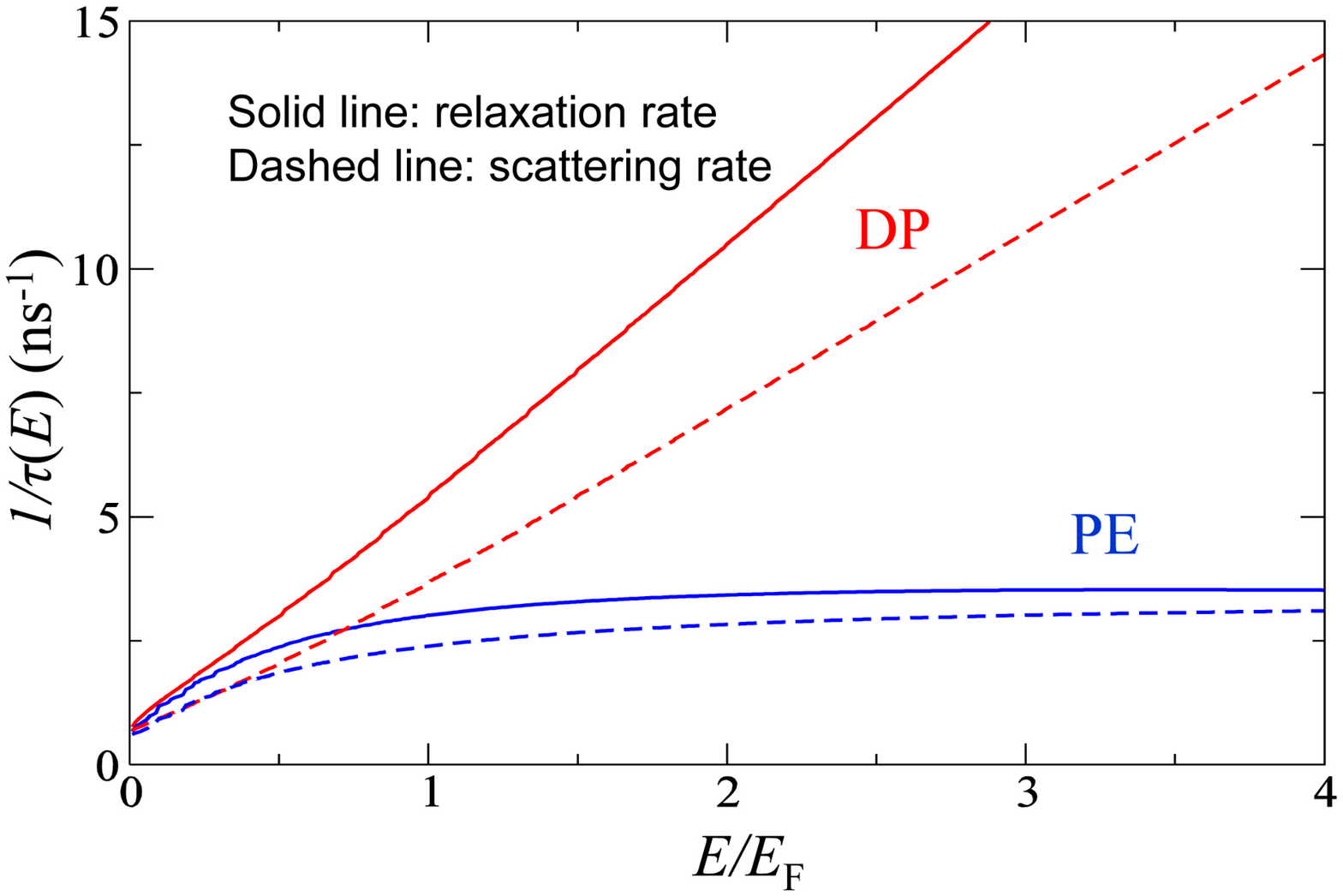}
\includegraphics[width=\linewidth]{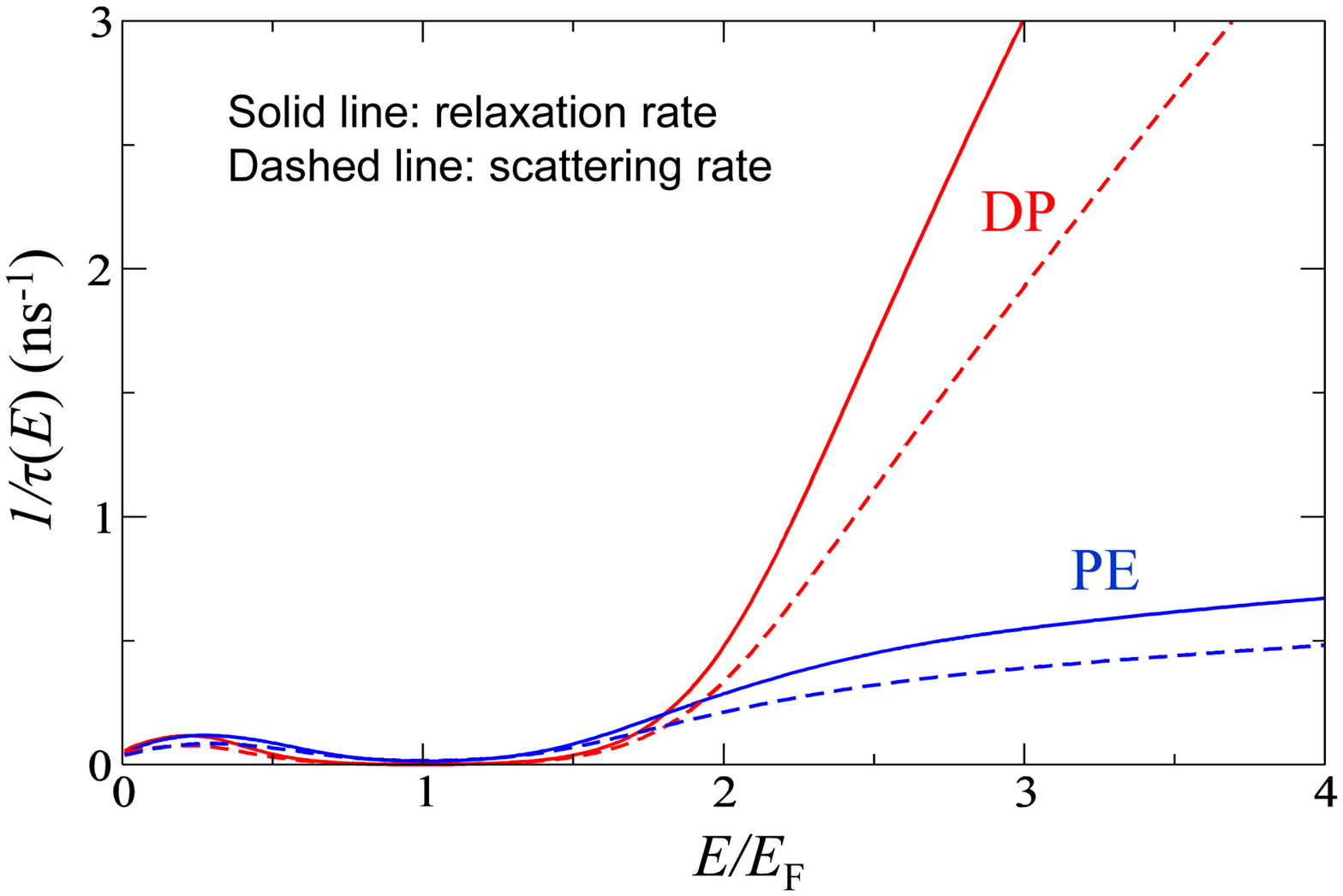}
\caption{(color online). The relaxation rate (solid line) and
  scattering rate (dashed line) as a function of hole energy $E$ for
  (a) $T=10$ K and (b) $T=1$ K with $n=10^{11}$ cm$^{-2}$, $n_{\rm
    imp}=0$, $n_{\rm depl}=0$ and $D=12.7$ eV. DP and PE represent the
  deformation potential and piezoelectric potential contributions,
  respectively.}  
\label{fig:relaxation_rate}
\end{figure}

%%%%%%%%%%%%%%%%%%%%%%%%%%%%%%%%%%%%%%%%%

\subsection{Acoustic phonon-limited transport}

Using the theoretical model outlined in Sec. III we study hole
transport limited by acoustic phonons in this subsection.
In Fig.~\ref{fig:relaxation_rate} we show the calculated 
scattering rates $\tau_{\rm s}^{-1}$ and transport relaxation rates
$\tau_{\rm t}^{-1}$ due to acoustic phonons, as a function of the hole energy. 
The relevant transport relaxation rate, $\tau_{\rm t}^{-1}$, has been obtained in Eq.~(\ref{eq:tau_QE}).
The two characteristic times shown in Fig.~\ref{fig:relaxation_rate} 
differ by the important $(1-\cos\theta)$ factor
\cite{dassarma_1985}. The scattering rate 
$\tau_{\rm s}^{-1}$ is given by making the replacement $(1-\cos\theta)
\rightarrow 1$ in the integrand for the formula for $\tau_{\rm t}^{-1}$
given in Eq.~(\ref{eq:tau_QE}). 
$\tau_{\rm t}$ determines the conductivity (or mobility),
$\sigma=ne\mu =ne^2\tau_{\rm t}/m$, where $n$ is the carrier density and
$\mu$ is the mobility, whereas $\tau_{\rm s}$ determines the quantum level
broadening, $\gamma =\hbar/2\tau_{\rm s}$, of the momentum eigenstates.
The scattering time $\tau_{\rm s}$ is related to the imaginary part of the
single-particle self-energy and simply gives the time 
between scattering events between a hole and an acoustic phonon.
The difference between $\tau_{\rm t}$ and $\tau_{\rm s}$ arises from the subtle effect
of the wave vector dependent transition rate \cite{dassarma_1985}.
The large angle scattering events (or large momentum transfer)
contribute significantly to the transport scattering events, 
but small angle scattering events 
where $\cos\theta \approx 1$ makes a negligible
contribution to $\tau_{\rm t}$, while all scattering events contribute
equally to $\tau_{\rm s}$. 
Our result for the individual DP and total PE rates are given in
Fig.~\ref{fig:relaxation_rate} for $n=10^{11}$ cm$^{-2}$ at two different
temperatures $T=10$ K and $T=1$ K. In this calculation we take $D=12.7$ eV
which is the best fitted values of the experimental data. 
We find that $\tau_{\rm t}/\tau_{\rm s} \approx 1$, since the screened 
electron-acoustic mode phonon interactions are of relatively short range.
It has been known that the ratio $\tau_{\rm t}/\tau_{\rm s}$ from remote ionized
impurities is much bigger due to the long-range nature of the electron-impurity
interaction \cite{harrang_1985}.

%%%%%%%%%%%%%%%%%%%%%%%%%%%  Fig. 5 %%%%%%%%%%%%%%%%%%%%%%
\begin{figure}%[htb]
\includegraphics[width=\linewidth]{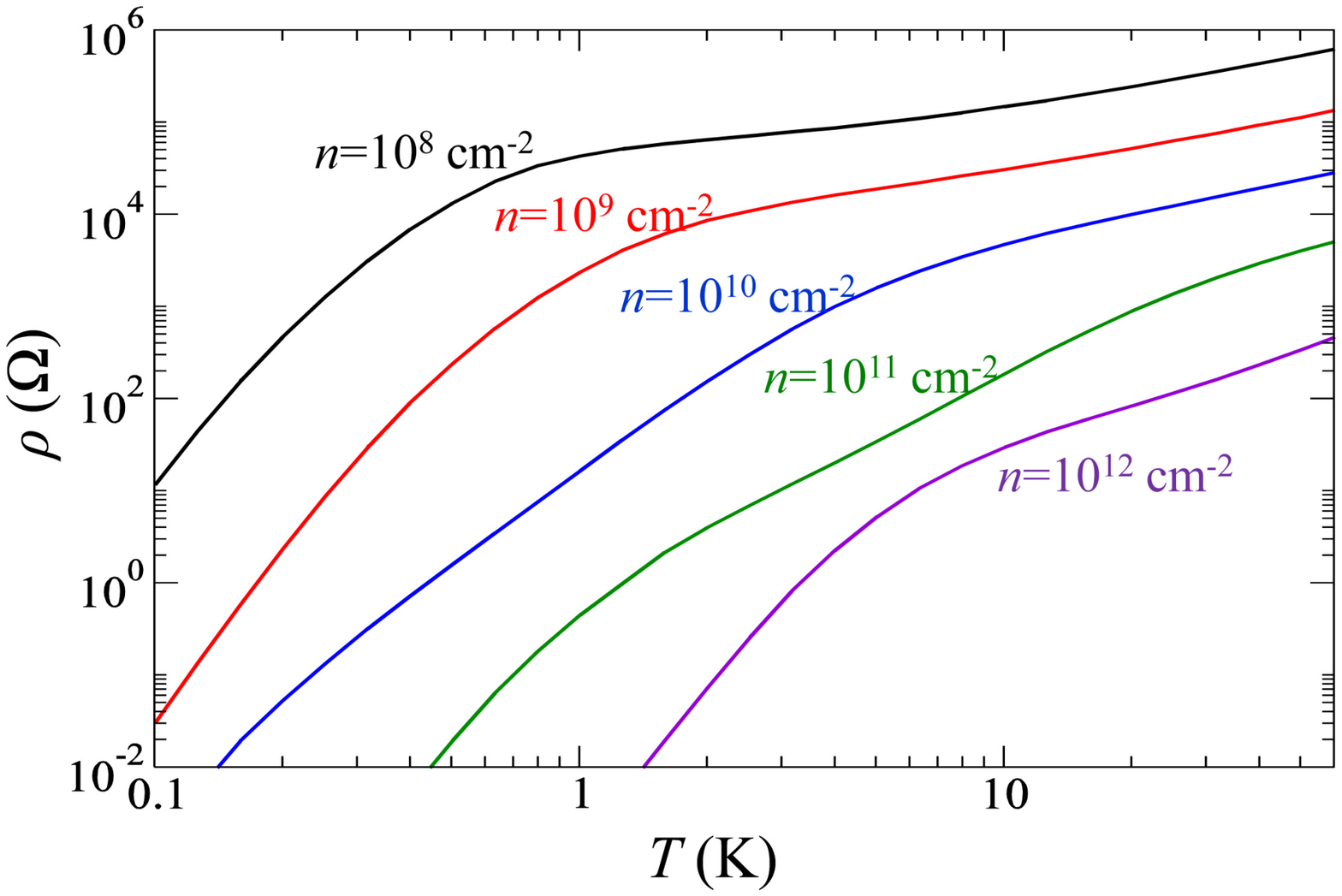}
\includegraphics[width=\linewidth]{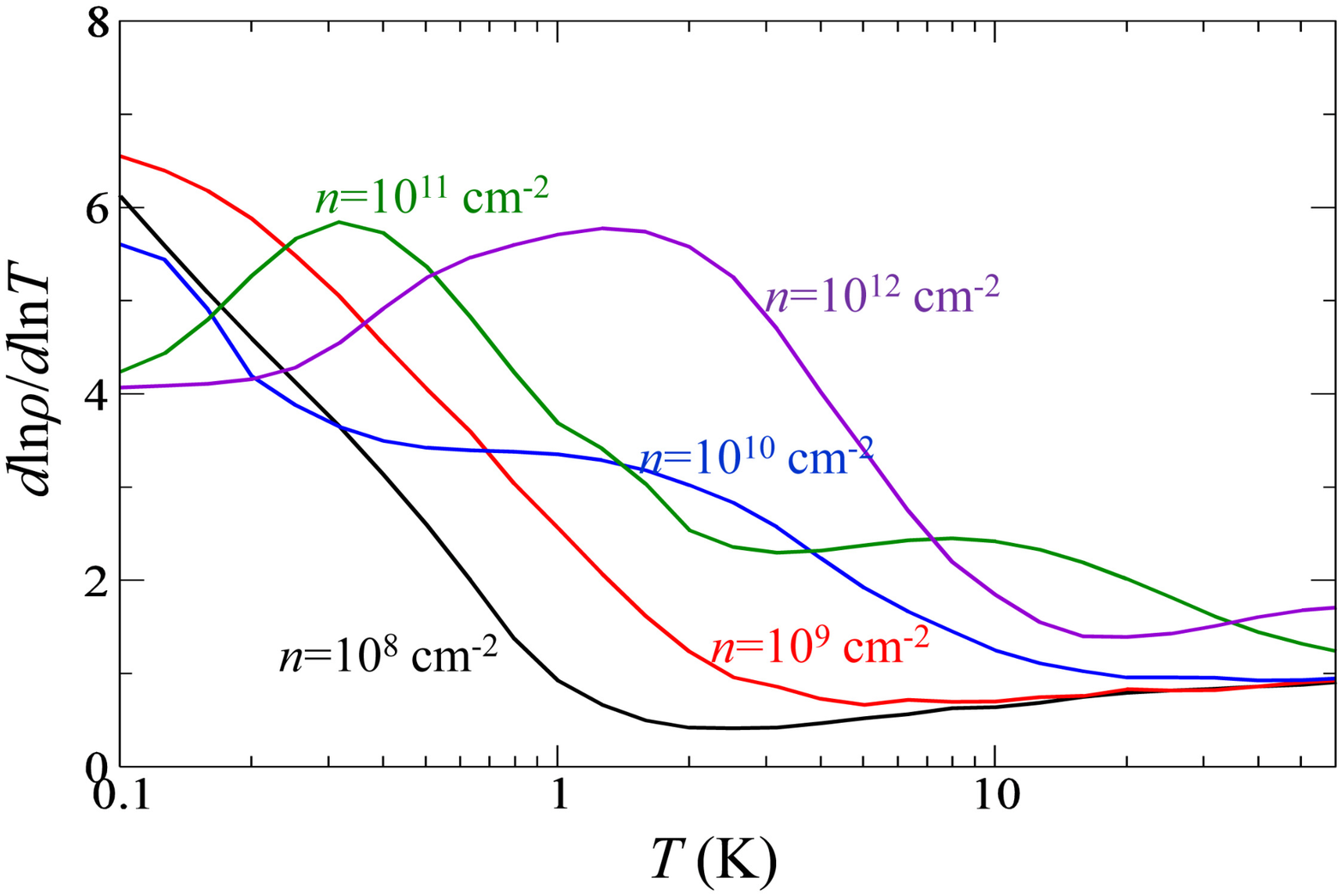}
\caption{(color online). (a) Acoustic phonon-limited resistivity of 2DHS as a function
  of temperature for $n=10^8$, $10^9$, $10^{10}$, $10^{11}$, $10^{12}$
  cm$^{-2}$ and (b) the calculated exponent $a$ in $\rho(T)
  \propto T^a$ which is obtained from logarithmic derivatives of (a).} 
\label{fig:rho_T_n}
\end{figure}

Figure \ref{fig:rho_T_n}(a) shows acoustic phonon-limited resistivity of
2DHS in the absence of impurity scattering as a function of temperature for
different hole densities $n=10^8$, $10^9$, $10^{10}$, $10^{11}$,
$10^{12}$ cm$^{-2}$. The calculated resistivities clearly demonstrate
the two different regimes: BG region characterized by high power law
behavior for $T < T_{\rm BG}$ and equipartition region with $\rho \sim T$ behavior
for $T > T_{\rm BG}$. The transition temperature $T_{\rm BG}$ increases with
density since $T_{\rm BG}\propto \sqrt{n}$.  As the density increases the calculated
resistivity at a fixed temperature decreases. 
In Fig.~\ref{fig:rho_T_n}(b) we show the logarithmic derivatives of
the acoustic phonon limited resistivity, which give rise to an
approximate temperature exponent of acoustic phonon limited
resistivity by writing  $\rho \sim T^a$, i.e., $a = d \ln \rho/ d \ln T$. 
At low temperature BG regime $T < T_{\rm BG}$ the numerically evaluated  
exponent $a$ varies from 4 to 6 depending on the carrier density. 
But at high temperatures the exponent approaches to 1 as we
expected, i.e., $\rho(T)\propto T$.

%%%%%%%%%%%%%%%%%%%%%%%%%% Fig. 6 %%%%%%%%%%%%%%%%%%%%%%%
\begin{figure}%[htb]
\includegraphics[width=\linewidth]{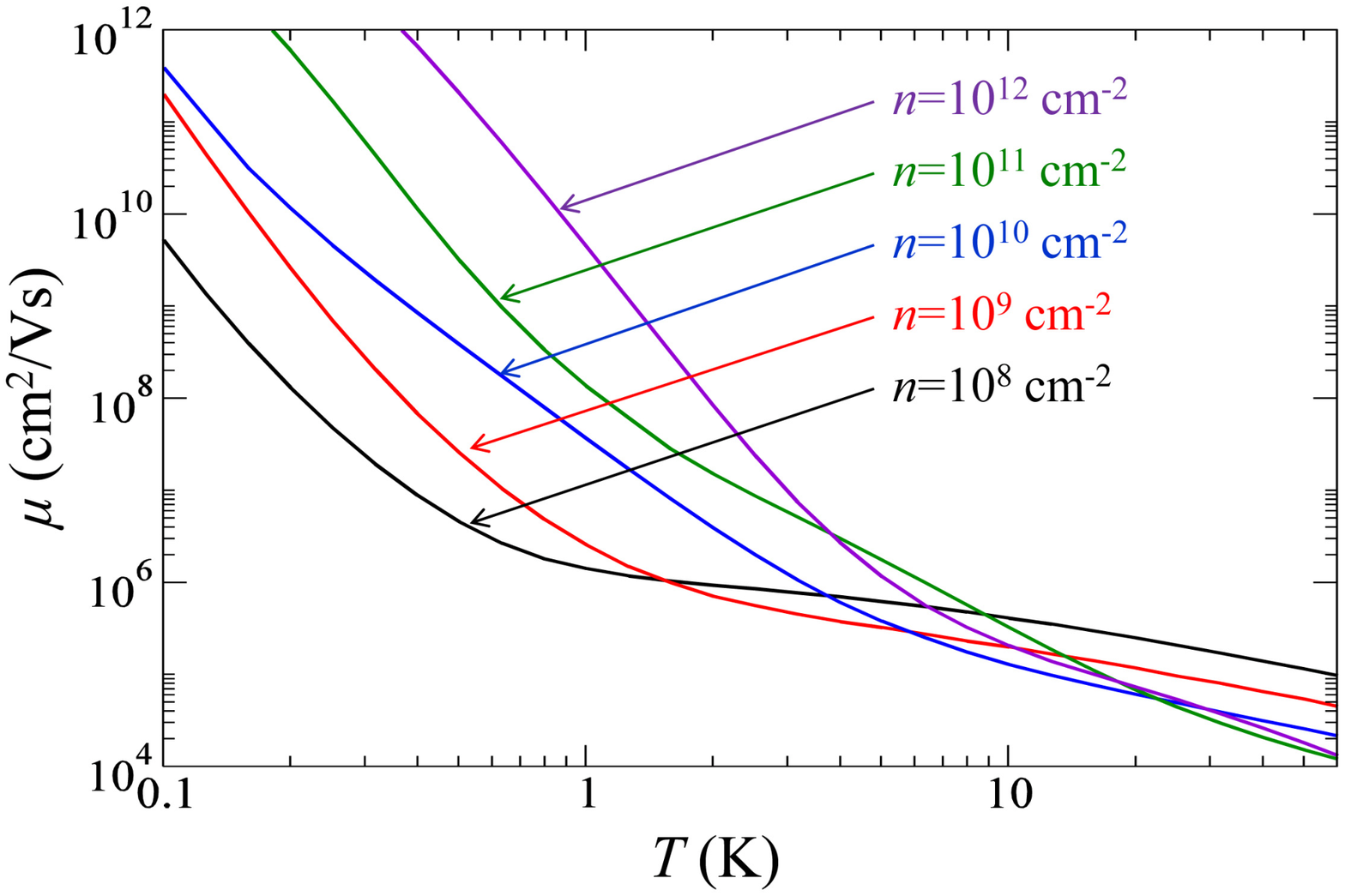}
\includegraphics[width=\linewidth]{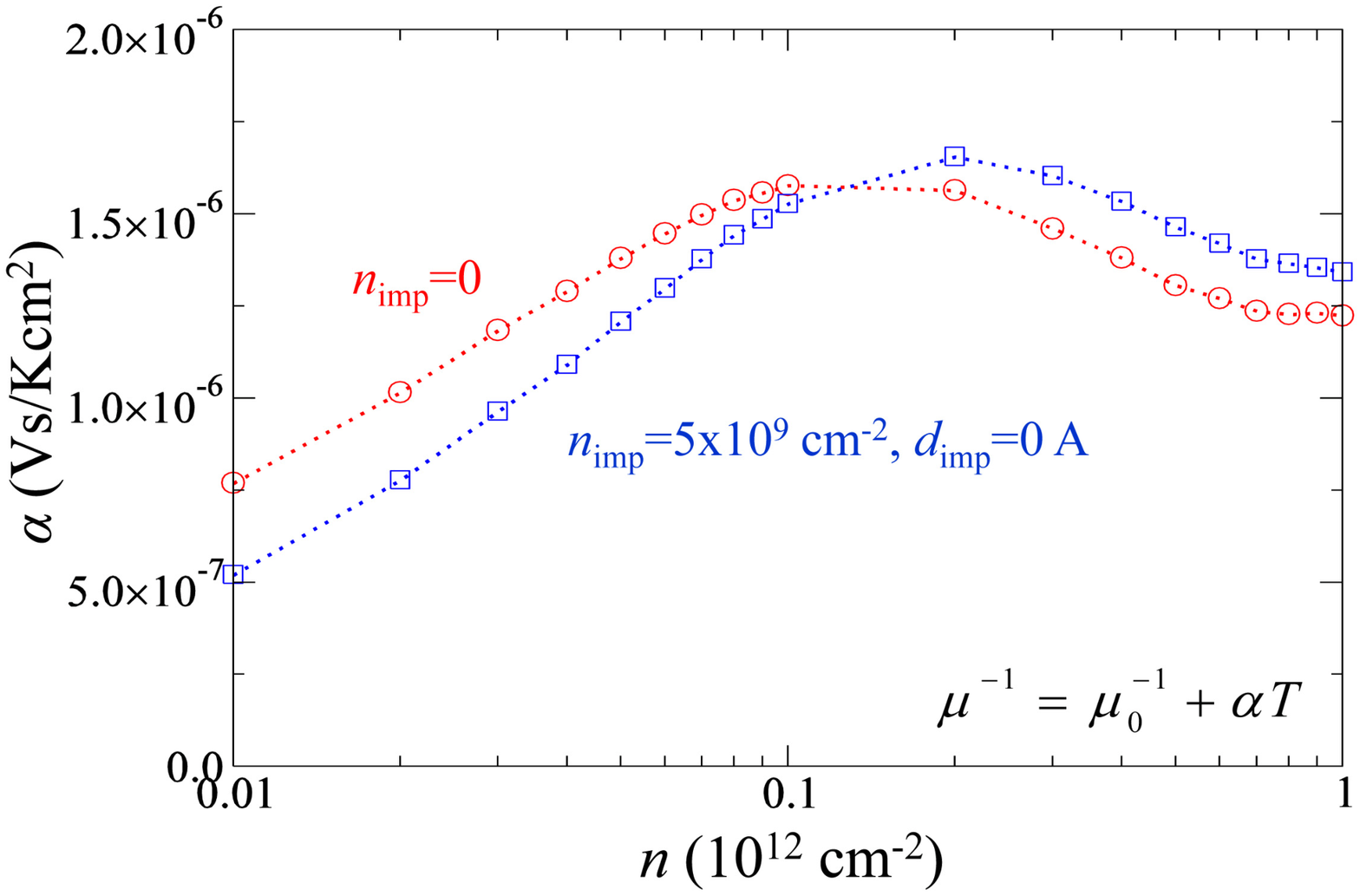}
\caption{(color online). (a) Acoustic phonon-limited mobility of 2DHS as a function of
  temperature for $n=10^8$, $10^9$, $10^{10}$, $10^{11}$, $10^{12}$
  cm$^{-2}$. (b) Density dependence of coefficient $\alpha$ for
  $n_{\rm imp}=$0 and 5$\times 10^9$ cm$^{-2}$, where
  $1/\mu=1/\mu_0+\alpha T$ in 10 K$<T<$60 K range.} 
\label{fig:mu_T}
\end{figure}

Figure \ref{fig:mu_T}(a) shows acoustic phonon-limited mobility of
2DHS in the absence of impurity as a function of temperature for
different hole densities
$n=10^8$, $10^9$, $10^{10}$, $10^{11}$, $10^{12}$ cm$^{-2}$. At high
temperatures ($T>10$ K) the calculated mobilities show very weak
density dependence for the density range $n=10^{10}-10^{12}$ cm$^{-2}$ and
decrease approximately as $\mu \sim T^{-1}$. Thus, 
the reciprocal mobility increases linearly with temperature,
i.e., $1/\mu=1/\mu_0+\alpha T$, where $\alpha$ is the slope in the
relation between $\mu^{-1}$ and $T$. 
Fig.~\ref{fig:mu_T}(b) shows density dependence of the slope $\alpha$ for
$n_{\rm imp}=$0 and 5$\times 10^9$ cm$^{-2}$ in the temperature range
10 K$<T<$60 K. The slope $\alpha$ first increases with $n$, reaches its
maximum at $n\sim 10^{11}$ cm$^{-2}$, and decreases very slowly for $n \agt 10^{11}$ cm$^{-2}$. 
This nonmonotonic behavior is different from
that of the n-type GaAs, in which the slope $\alpha$ has a minimum
value rather than a maximum \cite{kawamura1992,harris1990}.

%%%%%%%%%%%%%%%%%%%%%%%%%% Fig. 7 %%%%%%%%%%%%%%%%%%%%%%%
\begin{figure}%[htb]
\includegraphics[width=\linewidth]{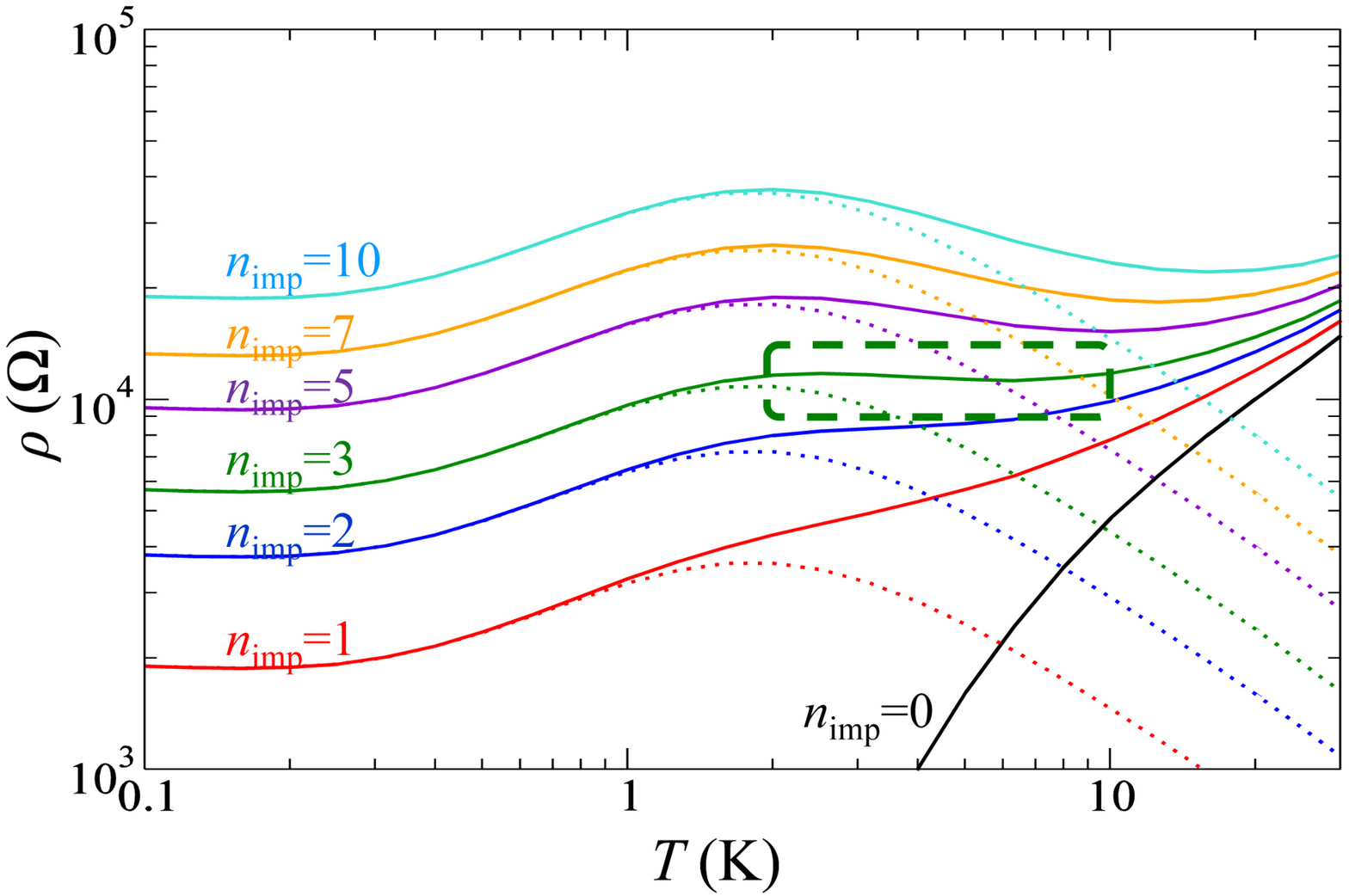}
\includegraphics[width=\linewidth]{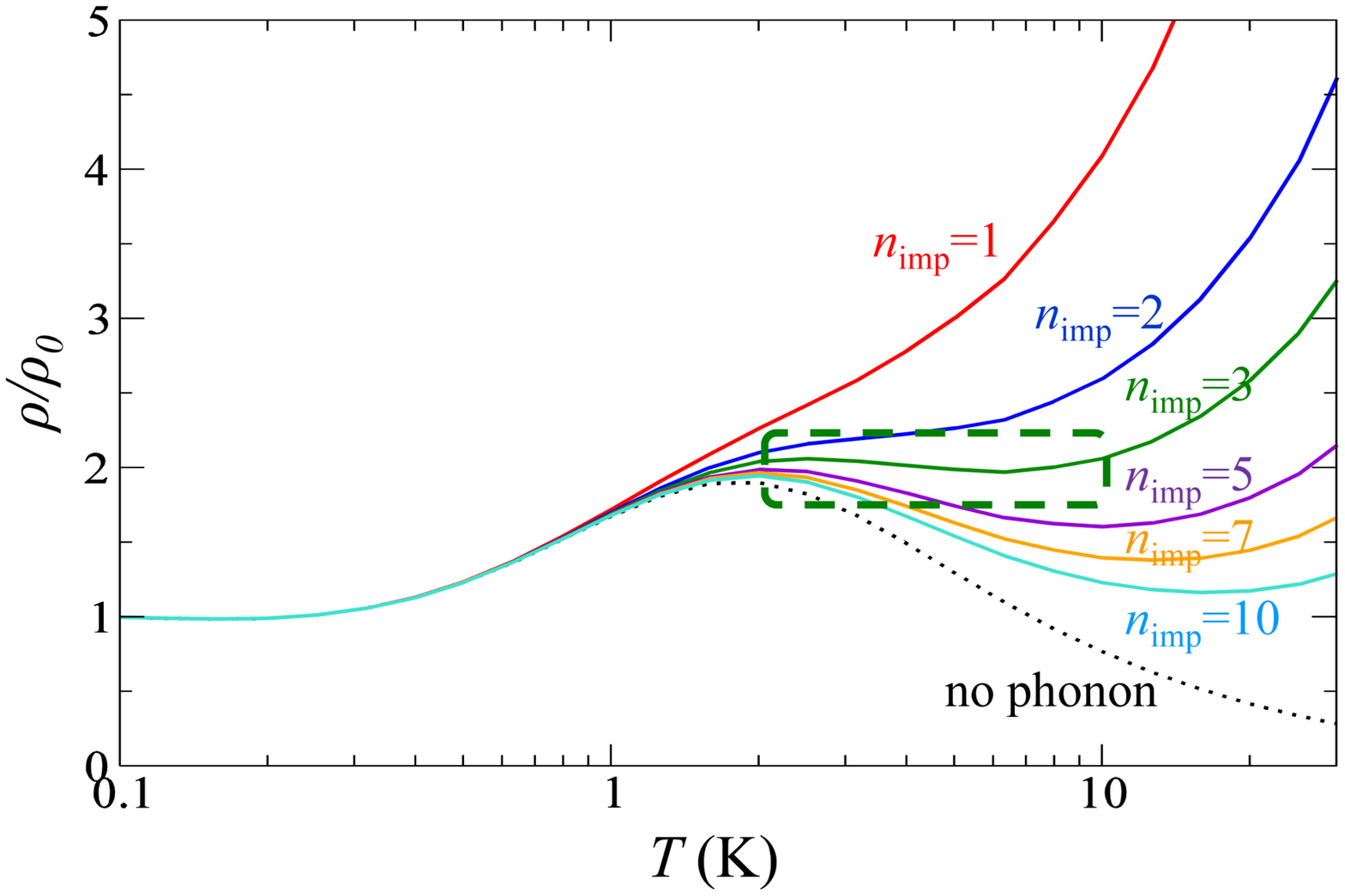}
\caption{(color online). (a) Resistivity of 2DHS as a function of temperature for
  $n_{\rm imp}=$0, 1, 2, 3, 5, 7, 10$\times 10^9$ cm$^{-2}$
 with $n=10^{10}$ cm$^{-2}$, $n_{\rm depl}=0$ and $d_{\rm imp}=0$. Dotted lines indicate the calculated
  resistivity due to the charged impurity scattering alone. (b) Same as (a) but rescaled by
  $\rho_0=\rho(T=0.1$ K).} 
\label{fig:rho_T_ni}
\end{figure}
%%%%%%%%%%%%%%%%%%%%%%%%%%%%%%%%%%%%%%%%%%%%%%%%%%%%%%%

%%%%%%%%%%%%%%%%%%%%%%%%%% Fig. 8 %%%%%%%%%%%%%%%%%%%%%%%%
\begin{figure}%[htb]
\includegraphics[width=\linewidth]{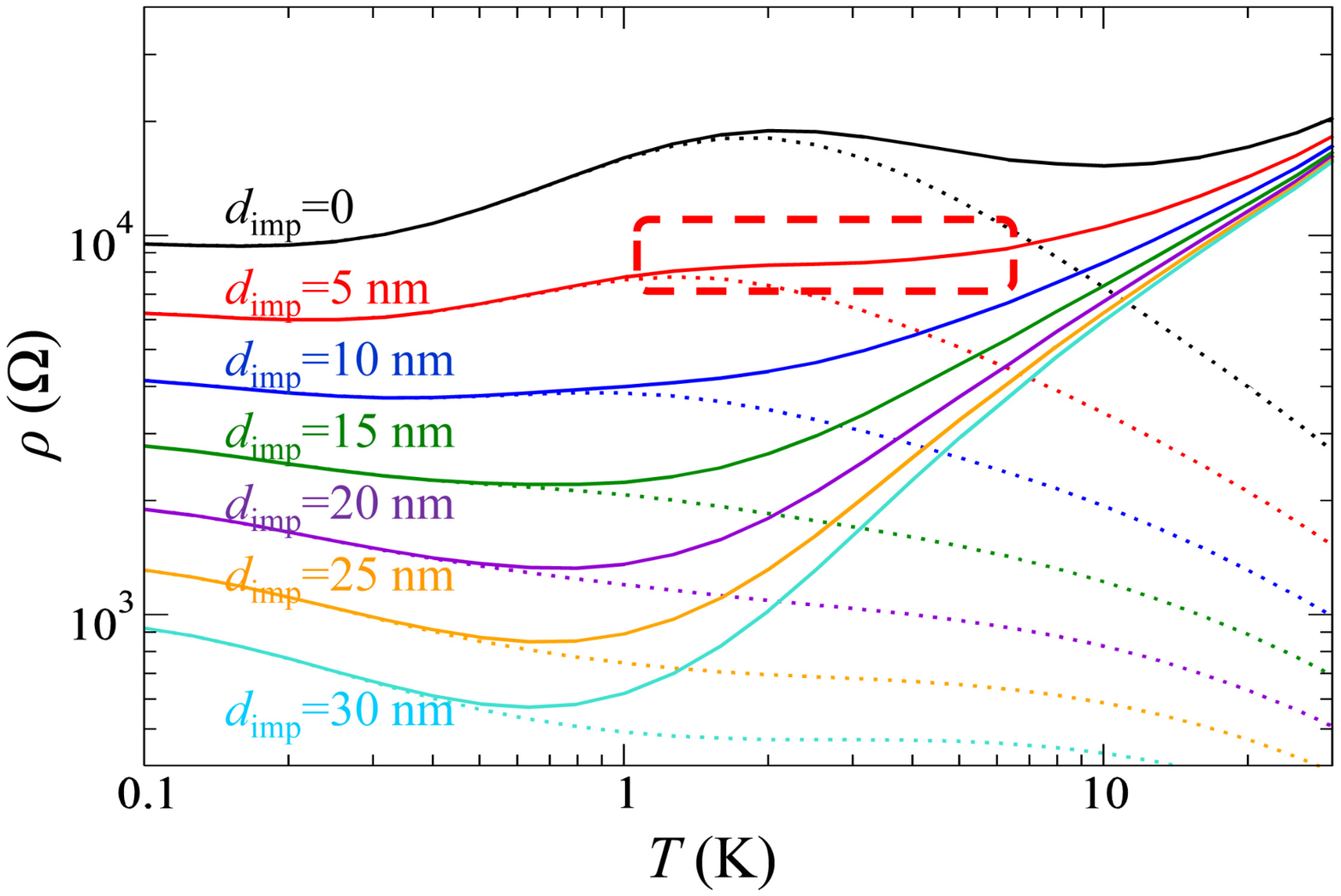}
\includegraphics[width=\linewidth]{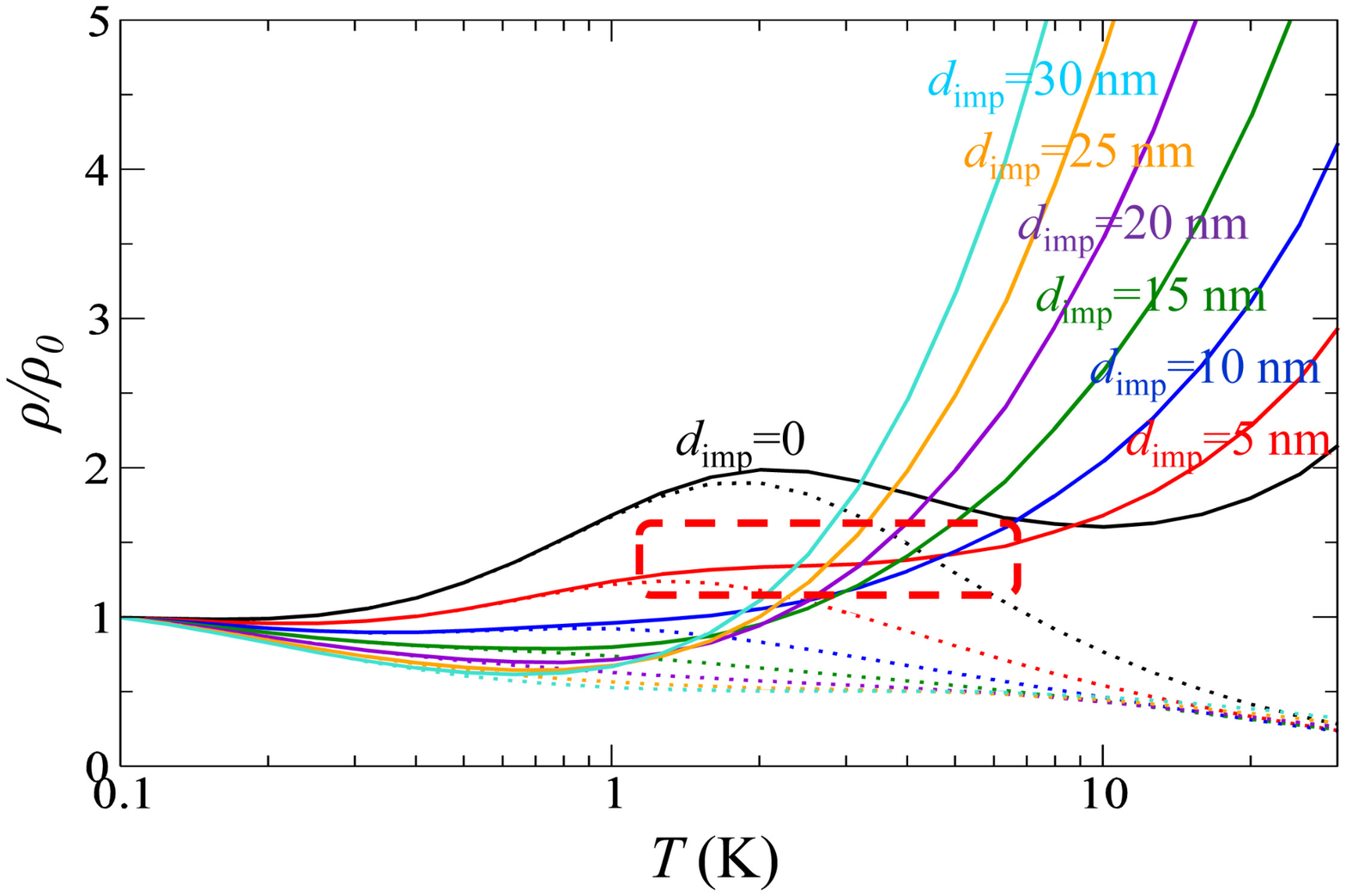}
\caption{(color online). (a) Resistivity of 2DHS as a function of temperature for
  $d_{\rm imp}=0, 5, 10, 15, 20, 25, 30$ nm with 
  $n=10^{10}$ cm$^{-2}$, $n_{\rm imp}=5\times 10^9$ cm$^{-2}$ and $n_{\rm 
    depl}=0$. Dotted lines indicate the calculated   resistivity due
  to the charged impurity scattering alone. (b) Same as (a) but
  rescaled by 
  $\rho_0=\rho(T=0.1$ K).} 
\label{fig:rho_T_di}
\end{figure}

\subsection{Nonmonotonic resistivity in p-GaAs}

In this subsection we study the transport in the presence of both
acoustic phonon and impurity scatterings and the nonmonotonic behavior
in temperature due to the competition between these two scatterings.

In Figs.~\ref{fig:rho_T_ni}  and \ref{fig:rho_T_di} we show our
calculated total resistivity $\rho(T)$ 
arising from screened charged impurity scattering $\rho_{\rm imp}(T)$ and
phonon scattering $\rho_{\rm ph}(T)$ as a function of temperature.
In Fig.~\ref{fig:rho_T_ni}(a) the total resistivity $\rho(T)$ is shown for
different impurity densities $n_{\rm imp}=$0, 1, 2, 3, 5, 7, 10$\times 10^9$ cm$^{-2}$ with a fixed $d_{\rm imp}=0$. 
%Figure \ref{fig:rho_T_ni}(b) is the same as Fig.~\ref{fig:rho_T_ni}(a) but rescaled by $\rho_0=\rho(T=0.1$ K). 
In Fig.~\ref{fig:rho_T_di}(a) the total resistivity $\rho(T)$ is shown for
different impurity location from the interface
$d_{\rm imp}=$0, 5, 10, 15, 20, 25, 30 nm with a fixed impurity density
$n_{\rm imp}=5\times 10^9$ cm$^{-2}$. 
Figures \ref{fig:rho_T_ni}(b) and \ref{fig:rho_T_di}(b) are the same as
Figs.~\ref{fig:rho_T_ni}(a) and \ref{fig:rho_T_di}(a), respectively,  but 
rescaled by $\rho_0=\rho(T=0.1$ K). 
In a real system the amount of random disorder depends on 
the strength and the spatial distribution of all the
random impurity scattering centers. However,
in these calculations we assume that the charged impurities
are randomly distributed in a 2D plane located at $d_{\rm imp}$ from the interface.
The calculation is carried out with the hole density
$n=10^{10}$ cm$^{-2}$ which corresponds to the Fermi
temperature $T_{\rm F}\approx 0.7$ K. The dotted lines in
Figs.~\ref{fig:rho_T_ni} and \ref{fig:rho_T_di} indicate the
calculated resistivity due to 
the charged impurity scattering alone, $\rho_{\rm imp}(T)$. To calculate
the total resistivity $\rho(T)$ we use the total scattering rate of Eq.~(\ref{eq:relaxation_time}) because 
the Matthiessen’s rule, which is implicitly assumed
$\rho(T)=\rho_{\rm imp}(T) + \rho_{\rm ph}(T)$, is known to be not strictly
valid at finite temperatures.

As shown in Fig.~\ref{fig:rho_T_ni}, when the charged impurity density $n_{\rm imp}$
increases the impurity scattering effects  
become stronger, while the phonon scattering effects are
unaffected. Therefore at high impurity densities the impurity
scatterings are dominant over phonon scatterings.
The calculated $\rho(T)$ increases at lower temperatures ($T<1$ K)
due to screening effects, then the quantum-classical crossover occurs
at the intermediate temperature regime around $T \sim 1.5$ K where nondegeneracy
effects make resistivity decrease as $\rho \sim T^{-1}$.
At higher temperatures ($T \gg 10$ K) phonon
scattering takes over and $\rho(T)$ increases with $T$. 
Thus, for large impurity densities
the temperature dependence of the calculated resistivity shows a
nontrivial nonmonotonic behavior, arising
from a competition among three mechanisms discussed above, i.e. screening which is
particularly important for $T<1$ K, nondegeneracy and the associated
quantum-classical crossover for $T \sim T_{\rm F}$, and 
phonon scattering effect which becomes increasingly important for $T > 10$ K. 
At lower impurity densities  the
quantum-classical crossover effects are not particularly shown in
Fig.~\ref{fig:rho_T_ni} 
because phonon scattering becomes more important than the classical
behavior $\rho \sim T^{-1}$, and the system makes a transition
from the quantum regime to the phonon scattering dominated regime. The
linear rise in $\rho(T)$ for $T>10$ K in Fig.~\ref{fig:rho_T_ni} is the
phonon scattering effect.

The same results shown in Fig.~\ref{fig:rho_T_ni} are expected by varying the
impurity location because the scattering limited by the remote impurity
becomes weaker as the distance of the impurity from the interface increases.  
In Fig.~\ref{fig:rho_T_di} we show the several different kinds of
nonmonotonic behavior by varying the impurity location. When the
impurities are located very close to the interface (top lines in
Fig.~\ref{fig:rho_T_di}) the nonmonotonic
behavior of the resistivity clearly appears in the temperature range
we consider (i.e. $T<100$ K) due to competition among the three
mechanisms discussed above. As the separation increases the
nonmonotonicity becomes weaker because of
the reduction of the charged impurity scattering and the associated
weakening of screening effects. In addition, the increase of
the separation gives rise to  
the shift of the local maximum peak to the lower temperature. 
For large separations (bottom lines in Fig.~\ref{fig:rho_T_di})
the local maximum peak does not appear in the calculated resistivity 
because it shifts to very low temperatures  ($T<0.1$ K).

One interesting finding in our calculation is the temperature region
where the calculated resistivity is approximately constant, as indicated 
by the dashed box in Figs.~\ref{fig:rho_T_ni} and \ref{fig:rho_T_di}.
The temperature range of the constant resistivity appears when the
increasing resistivity 
due to phonon scatterings compensates for the decreasing resistivity
due to the nondegeneracy effects. The flat region depends critically on the
impurity density and the location of the impurities, and can be
observed in experiments by varying the doping density and location. In 
Fig.~\ref{fig:rho_T_ni} a flat region spanning around 2 K $< T <$ 10 K 
appears at an impurity density $n_{\rm imp} = 3\times 10^9$ cm$^{-2}$
for $d_{\rm imp}=0$. In 
Fig.~\ref{fig:rho_T_di} the flat region for 2 K $< T <$ 10 K 
appears at $d_{\rm imp}=5$ nm with an impurity density $n_{\rm imp} =
5\times 10^9$ cm$^{-2}$.
It is, therefore, possible in some situations for a complete
accidental cancellation between the increasing temperature dependence
of the phonon-induced resistivity and the decreasing temperature
dependence of the quantum classical crossover effect from impurity
scattering in a narrow intermediate temperature regime. We believe
that this has recently been observed experimentally \cite{zhou2012}, but a
detailed comparison with experiment is not possible due to the
complications of the parallel magnetic field used in the experimental
measurement to induce magneto-orbital coupling.

\section{Conclusion}
\label{sec:discussion}

To conclude, 
we have calculated the temperature dependent transport
properties of p-type GaAs-based 2DHSs for temperatures
$T \alt 100$ K by taking into account both hole-phonon and
hole-impurity scatterings. Our theory includes
temperature-dependent screening of both charged impurity scattering and phonon
scattering. We extract the deformation potential $D$ of hole-phonon
coupling constant by fitting the experimentally available mobility data.
We find that the deformation potential coupling
varies (i.e. $D=7.6-12.7$ eV) depending on the
value of the depletion density $n_{\rm depl}$, which is not known.
When we assume $n_{\rm depl}=0$ we obtain $D=12.7$ eV for the p-GaAs
acoustic phonon deformation potential,  
which is larger than the generally accepted value in bulk GaAs ($D=7$
eV) \cite{wolfe1970} but comparable to the value of the
n-GaAs ($12-14$ eV) \cite{price1985,mendez1984,kawamura1990} in 2D
electron systems.

We also investigate the nonmonotonicity of $\rho(T)$  arising
from the competition among three mechanisms: screening, nondegeneracy,
and phonon scattering. Both 
screening and phonon scattering	mechanisms give rise to
monotonically increasing $\rho(T)$ with $T$ (at low temperature for screening, and
at high temperatures for phonons), but nondegeneracy effects produce a
$\rho(T)$ decreasing with increasing $T$ for $T > T_{\rm F}$. Since
phonon scattering is the dominant temperature-dependent scattering
mechanism in GaAs holes	for $T \agt 5-10$ K, depending on the density,
the stronger 
nonmonotonicity appears when the impurity scattering is dominant
over phonon scattering below $T\sim 5-10$ K. 
We carefully study the non-trivial transport properties of p-GaAs at the
intermediate temperature range (i.e., 2 K$<T<$10 K). Interestingly we
find that the approximate temperature independence may appear
in which $\rho(T)$ saturates in an intermediate temperature range,
arising from the approximate
cancellation between the quantum-classical crossover and phonon
scattering. Since this flat region of the temperature dependent
resistivity depends critically on the
impurity density and the location of the impurities, it can be
observed in experiments by varying the doping density and location.
We believe that a recent measurement \cite{zhou2012} has observed this
saturation effect.

\section*{acknowledgments}

The work is supported by the NRI-SWAN and US-ONR. We thank
Dr. H. Noh for sharing unpublished data with us. 

%%%%%%%%%%%%%%%%%%%%%%%%%%%%%%%%%%%%%%%%%%%%%%%%%%


\begin{thebibliography}{999}

\bibitem{ziman} J. M. Ziman, {\it Electrons and Phonons} (Oxford University Press, New York, 1963). 

\bibitem{g_phonon1}E. H. Hwang and S. Das Sarma, \prb {\bf 77}, 115449 (2008); 
Hongki Min, E. H. Hwang, and S. Das Sarma, \prb {\bf 83}, 161404 (2011); 
D. K. Efetov and Philip Kim, Phys. Rev. Lett. {\bf 105}, 256805 (2010). 

\bibitem{holeexp} Y. Hanein, U. Meirav, D. Shahar, C. C. Li, D. C. Tsui, and H. Shtrikman, Phys. Rev. Lett. {\bf 80}, 1288 (1998); 
M. Y. Simmons, A. R. Hamilton, M. Pepper, E. H. Linfield, P. D. Rose, and D. A. Ritchie, Phys. Rev. Lett. {\bf 80}, 1292 (1998);
J. Yoon, C. C. Li, D. Shahar, D. C. Tsui, and M. Shayegan, Phys. Rev. Lett. {\bf 82}, 1744 (1999);
M. J. Manfra, E. H. Hwang, S. Das Sarma, L. N. Pfeiffer, K. W. West, and A. M. Sergent, Phys. Rev. Lett. 99, 236402 (2007).  

\bibitem{mills1999} A. P. Mills, Jr., A. P. Ramirez, L. N. Pfeiffer, and K. W. West, Phys. Rev. Lett. {\bf 83}, 2805 (1999). 

\bibitem{kawamura1992} T. Kawamura and S. Das Sarma, Phys. Rev. B {\bf 45}, 3612 (1992).  
%Phonon-scattering-limited electron mobilities in AlxGa1-xAs/GaAs
%heterojunctions

\bibitem{zhou2012} X. Zhou, B. Schmidt, L. W. Engel, G. Gervais,  L. N. Pfeiffer, K. W. West, and S. Das Sarma, Phys. Rev. B {\bf 85},
  041310 (2012). 

\bibitem{parallel_B}
X. Zhou, B. A. Piot, M. Bonin, L. W. Engel, S. Das Sarma, G. Gervais,
L. N. Pfeiffer, and K. W. West, Phys. Rev. Lett. {\bf 104}, 216801 (2010);
%Colossal Magnetoresistance in an Ultraclean Weakly Interacting 2D Fermi Liquid
S. Das Sarma and E. H. Hwang, Phys. Rev. Lett. {\bf 84}, 5596 (2000).
%Parallel Magnetic Field Induced Giant Magnetoresistance in Low Density Quasi-Two-Dimensional Layers


\bibitem{kawamura1990}
T. Kawamura and S. Das Sarma, Phys. Rev. B {\bf 42}, 3725 (1990). 

\bibitem{hwang2000} S. Das Sarma and E. H. Hwang, Phys. Rev. B {\bf 61}, R7838 (2000). 

%\bibitem{kravchenko1994} S. V. Kravchenko, G. V. Kravchenko,
%  J. E. Furneaux, V. M. Pudalov, and M. D'Iorio, 
%  Phys. Rev. B {\bf 50}, 8039 (1994).

\bibitem{dassarma2005} S. Das Sarma and E. H. Hwang, Solid  State Commun. {\bf 135}, 579 (2005).

\bibitem{abrahams2001} E. Abrahams, S. V. Kravchenko, and M. P. Sarachik, Rev. Mod. Phys. {\bf 73}, 251 (2001).

%\bibitem{spivak2010} B. Spivak, S. V. Kravchenko, S. A. Kivelson, and X. P. A. Gao, Rev. Mod. Phys. {\bf 82}, 1743 (2010).

\bibitem{dassarma1999} S. Das Sarma and E. H. Hwang, \prl {\bf 83} 164 (1999).

\bibitem{zna} G. Zala, B. N. Narozhny, and I. L. Aleiner, Phys. Rev. B {\bf 64}, 214204 (2001).

\bibitem{noh2003} Hwayong Noh, M. P. Lilly, D. C. Tsui, J. A. Simmons, E. H. Hwang, S. Das Sarma, L. N. Pfeiffer, and K. W. West, Phys. Rev. B 68, 165308 (2003); Hwayong Noh {\it et al.}, unpublished.

\bibitem{wolfe1970} C. M. Wolfe, G. E. Stillman, and W. T. Lindley, J. Appl. Phys. {\bf 41}, 3088 (1970).

\bibitem{price1985} P. J. Price, \prb {\bf 32}, 2643 (1985).

\bibitem{mendez1984} E. E. Mendez, P. J. Price, and M. Heiblum, Appl. Phys. Lett. {\bf 45}, 294 (1984).

%\bibitem{dassarma2000} S. Das Sarma and E. H. Hwang, Phys. Rev. B {\bf 61}, R7838 (2000).

\bibitem{lilly2003} M. P. Lilly, J. L. Reno, J. A. Simmons,
  I. B. Spielman, J. P. Eisenstein, L. N. Pfeiffer, K. W. West,
  E. H. Hwang, and S. Das Sarma, Phys. Rev. Lett. {\bf 90}, 056806 (2003).

\bibitem{ando1982}
T. Ando,  A. B. Fowler, and F. Stern, Rev. Mod. Phys. {\bf 54}, 437 (1982).
%Electronic properties of two-dimensional systems

\bibitem{fang1966}
F. F. Fang and W. E. Howard, Phys. Rev. Lett. {\bf 16}, 797 (1966).
%Negative Field-Effect Mobility on (100) Si Surfaces

\bibitem{zook1964} J. D. Zook, Phys. Rev. A{\bf 136}, 869 (1964); P. J. Price, \prb {\bf 32}, 2643 (1985). 

\bibitem{dassarma2011} S. Das Sarma, S. Adam, E. H. Hwang, and E. Rossi, \rmp {\bf 83}, 407 (2011).

\bibitem{dassarma2004} S. Das Sarma and E. H. Hwang, \prb {\bf 69}, 195305 (2004);
%Metallicity and its low-temperature behavior in dilute two-dimensional carrier systems
%S. Das Sarma and E. H. Hwang, 
\prb {\bf 68}, 195315 (2003).
%Low-density finite-temperature apparent insulating phase in two-dimensional semiconductor systems

%\bibitem{heremans1994} J. J. Heremans, M. B. Santos, K. Hirakawa, and
%  M. Shayegan, J. Appl. Phys. {\bf 76}, 1980 (1994).
%Temperature dependence of the low-temperature mobility in ultrapure AlxGa1-xAs/GaAs heterojunctions: Acoustic-phonon scattering

\bibitem{stormer1984} H. L. St\"{o}rmer, A. C. Gossard, W. Wiegmann, R. Blondel, and K. Baldwin, Appl. Phys. Lett. {\bf 44}, 139 (1984);
E. E. Mendez and W. I. Wang, Appl. Phys. Lett. 46, 1159 (1985);
K. Oe and K. Tsubaki, J. Appl. Phys. 59, 3527 (1986).
%Temperature dependence of the mobility of two-dimensional hole systems in modulation-doped GaAs-(AlGa)As

%\bibitem{noh2003}
%H. Noh, M. P. Lilly, D. C. Tsui, J. A. Simmons, E. H. Hwang, S. Das Sarma, L. N. Pfeiffer, and K. W. West, Phys. Rev. B {\bf 68}, 165308 (2003).
%Interaction corrections to two-dimensional hole transport in the large-rs limit

%\bibitem{sarma2000}
%S. Das Sarma and E. H. Hwang, Phys. Rev. B {\bf 61}, 7838(R) (2000).
%Calculated temperature-dependent resistance in low-density two-dimensional hole gases in GaAs heterostructures

%\bibitem{mills1999}
%A. P. Mills, Jr., A. P. Ramirez, L. N. Pfeiffer, and K. W. West, Phys. Rev. Lett. {\bf 83}, 2805 (1999).
%Nonmonotonic Temperature-Dependent Resistance in Low Density 2D Hole Gases

\bibitem{dassarma_1985} S. Das Sarma and F. Stern, Phys. Rev. B {\bf 32},
8442 (1985); E. H. Hwang and S. Das Sarma, \prb {\bf 77}, 195412 (2008).


\bibitem{harrang_1985} J. P. Harrang, R. J. Higgins, R. K. Goodall, P. R. Jay, M. Laviron, and P. Delescluse, Phys. Rev. B {\bf 32}, 8126 (1985);
R. G. Mani and J. R. Anderson, {\it ibid.} {\bf 37}, 4299 (1988);
M. Sakowicz, J. Lusakowski, K. Karpierz, M. Grynberg, and B. Majkusiak, Appl. Phys. Lett. {\bf 90}, 172104 (2007). 

\bibitem{harris1990} J. J. Harris, C. T. Foxon, D. Hilton, J. Hewett, C. Roberts, and S. Auzoux, Surf. Sci. {\bf 229}, 113 (1990).

\end{thebibliography}
\end{document}